\documentclass[3p]{elsarticle}
\usepackage{epstopdf}
\usepackage{subfigure,subcaption,graphicx,float}
\usepackage{amsmath, amsfonts, amssymb}
\usepackage{amsthm}
\usepackage{yhmath}
\usepackage{epsf}
\usepackage{amsfonts}
\usepackage{stmaryrd}
\usepackage{amssymb}
\usepackage{leftidx}
\usepackage{xcolor}
\usepackage{mathtools}
\usepackage{placeins}
\usepackage{booktabs}
\usepackage{enumitem}
\usepackage{caption}
\usepackage{multirow}
\usepackage{stackengine}
\usepackage[utf8]{inputenc}
\usepackage[english]{babel}
\usepackage{bm}
\usepackage{array}
\usepackage{siunitx}
\usepackage{url}
\usepackage{algorithm}
\usepackage{algpseudocode}

\let\caronaccent\v   
\def\bfu{{\bf u}}

\def\bfy{{\bf y}}

\def\bfE{{\bf E}}

\def\bfI{{\bf I}}

\def\bfR{{\bf R}}
\def\bfS{{\bf S}}

\def\bfX{{\bf X}}
\def\bfF{{\bf F}}
\def\bfU{{\bf U}}

\def\eps{\varepsilon}

\def\e0{\varepsilon_0}

\def\s0{\sigma_0}
\def\v{\varphi}

\def\sts{\sigma_{\texttt{ts}}}
\def\scs{\sigma_{\texttt{cs}}}
\def\shs{\sigma_{\texttt{hs}}}
\def\sss{\sigma_{\texttt{ss}}}
\def\de{\delta^\varepsilon}
\def\ce{c_{\texttt{e}}}

\DeclareMathAlphabet{\mathsfit}{T1}{\sfdefault}{\mddefault}{\sldefault}
\SetMathAlphabet{\mathsfit}{bold}{T1}{\sfdefault}{\bfdefault}{\sldefault}
\let\v\caronaccent

\algrenewcommand\algorithmicrequire{\textbf{Input:}}
\algrenewcommand\algorithmicensure{\textbf{Output:}}

\usepackage{listings}
\usepackage{xcolor}

\newcommand{\R}{\mathbb{R}}
\newcommand{\hmin}{h_{\min}}
\newcommand{\Ibar}{\bar{\mathcal{I}}}
\newcommand{\Tcal}{\mathcal{T}}
\newcommand{\Acal}{\mathcal{A}}

\theoremstyle{plain}

\long\def\symbolfootnote[#1]#2{\begingroup%
\def\thefootnote{\fnsymbol{footnote}}\footnote[#1]{#2}\endgroup}

\begin{document}
\begin{frontmatter}

\title{Adaptive Mesh Coarsening for Efficient Phase-Field Fracture Simulations \vspace{0.1cm}}

\vspace{-0.1cm}

\author[GT]{Aarosh Dahal}
\ead{adahal8@gatech.edu}

\author[Vanderbilt]{Abhinav Gupta}
\ead{abhinav.gupta@vanderbilt.edu}

\author[GT,GTmech]{Aditya Kumar\corref{cor1}}
\ead{aditya.kumar@ce.gatech.edu}

\address[GT]{School of Civil and Environmental Engineering, Georgia Institute of Technology, Atlanta, GA 30332, USA \vspace{0.05cm}}

\address[GTmech]{George W. Woodruff School of Mechanical Engineering, Georgia Institute of Technology, Atlanta, GA 30332, USA \vspace{0.05cm}}

\address[Vanderbilt]{Department of Civil and Environmental Engineering, Vanderbilt University,  Nashville, TN 37235, USA}

\cortext[cor1]{Corresponding author}

\begin{abstract}
\vspace{-0.1cm}

Phase-field models provide a versatile framework for simulating fracture nucleation and propagation without explicit crack tracking. Their principal computational challenge is the need to resolve a small regularization length with a sufficiently fine finite element mesh. This requirement can render uniform-mesh simulations prohibitively expensive, particularly for three-dimensional problems, problems involving distributed crack nucleation, and soft nearly incompressible materials. Adaptive mesh refinement offers a natural means of reducing this cost, but existing approaches often rely on heuristic refinement indicators, are primarily designed for problems containing pre-existing cracks, and retain increasingly large refined regions as cracks grow in size.

This work presents an adaptive mesh coarsening framework along with adaptive mesh refinement for phase-field fracture. The coarsening strategy replaces the fractured region in the crack wake with a coarse crack band that preserves the essential mechanical behavior of a crack, making it necessary to have a refined mesh only in a small region near the tip of the growing crack. The method introduces a physics-based refinement indicator derived from the violation of the material strength surface, which is a necessary condition for fracture evolution. The indicator therefore robustly identifies regions where crack nucleation or propagation is imminent and can be applied across arbitrary materials, geometries, and loading conditions.  The framework is implemented in parallel within FEniCSx;  the supporting finite element codes are made available. Its generality and substantial computational benefit are demonstrated through benchmark problems involving crack propagation and nucleation under quasi-static and dynamic loading, for soft and hard materials, and for thermomechanical fracture.

\keyword{Phase-field fracture; Adaptive mesh refinement; Adaptive mesh coarsening; FEniCSx; Parallel computing; Strength surface}
\endkeyword

\end{abstract}
\end{frontmatter}

\section{Introduction}\label{Sec:Introduction}

The study of crack growth in materials presents an immense theoretical and computational challenge. Theoretically, in elastic brittle materials, a critical challenge is that Griffith’s law does not provide information about where the crack grows. Many empirical criteria such as the maximum hoop stress criterion or the maximum energy release rate criterion have been proposed to address this challenge \cite{ErdoganSih1963, sih1974strain, goldstein1974brittle, nuismer1975energyrelease, wu1978fracture}, but their fundamental status is not clear. To align the study of crack growth with Griffith’s original vision of energetic competition, the variational approach to fracture in brittle materials under quasi-static loading was proposed by Francfort and Marigo \cite{Francfort98}. The variational approach readily admits a numerical implementation with the standard finite element method (FEM) when a phase field is introduced to regularize the crack surface \cite{Bourdin00}. The phase field, $v \in [0,1]$, is a continuous field variable that transitions smoothly from intact ($v=1$) to fractured ($v=0$) material over a regularization width $\varepsilon$.

The resulting phase-field approach to fracture offers two principal advantages over traditional crack-growth algorithms based, for example, on linear elastic fracture mechanics (LEFM): (i) it eliminates the need to evaluate the energy-release rate or stress-intensity factors at the crack tip, and (ii) it requires no additional empirical criterion to determine the crack path. Instead, crack growth is obtained entirely through the solution of two coupled partial differential equations. This formulation is therefore readily extendable to three-dimensional problems and to the fracture of soft materials. Whether it consistently predicts crack paths in quantitative agreement with experiments remains under investigation \cite{WK2025}; nevertheless, its computational implementation is by now well established. Extensions to dynamic, ductile, and thermomechanical fracture are also computationally natural \cite{borden2012phase, borden2016phase, bourdin2014morphogenesis}, although important theoretical questions remain unresolved in each setting.

These theoretical and computational advantages come at a high cost. For the regularized phase-field model to $\Gamma$-converge to the corresponding sharp-crack description, the regularization length $\varepsilon$ must tend to zero and be adequately resolved by the finite element mesh. In practical computations, however, $\varepsilon$ need not be infinitesimal; it must instead be sufficiently small relative to the smallest characteristic structural length scale, $L$, typically $\varepsilon \sim 10^{-2}L$. The widely used AT$_1$ phase-field regularization \cite{Tanne18}, which has compact support, generally requires an element size of approximately $h \sim \varepsilon/5$ within the fracture process zone. Consequently, even a simple square domain discretized with a uniform mesh would require $\mathcal{O}((L/h)^2) \sim 2.5 \times 10^5$ elements. In three dimensions, this requirement increases to $\mathcal{O}((L/h)^3) \sim 1.25 \times 10^8$ elements, rendering even the simplest computations impractical and potentially far more expensive than, say, LEFM-based implementations using G/X-FEM \cite{duarte2020validation}.

The computational cost increases further for soft incompressible materials, which may require higher-order or non-conforming finite elements \cite{KFLP18}, and for ductile materials, for which the intrinsic fracture length scale can be extremely small \cite{talamini2021attaining}. Incorporating fracture nucleation into the phase-field formulation imposes an additional computational burden because it introduces a finite intrinsic length scale that must also be adequately resolved by the mesh. Numerous approaches have been proposed to model nucleation in brittle materials. One class of models identifies the phase-field regularization length with the intrinsic tensile fracture length scale \cite{pham2011gradient, Tanne18}. For materials such as glass and elastomers, this may require length scales of approximately $0.1$ mm or smaller \cite{KLP20, KRLP22}, making large structural-scale computations prohibitively expensive. Other approaches provide formulations in which nucleation predictions are independent of the regularization length \cite{KBFLP20, Wu18}; nevertheless, they generally still require the regularization length not to substantially exceed the intrinsic fracture length scale.

Adaptive mesh refinement (AMR) in the vicinity of a growing crack provides a natural means of reducing the computational cost of phase-field fracture simulations and has been studied extensively over the past decade. Although AMR has proven effective, most existing studies have focused on compressible, isotropic, linear-elastic materials and problems involving pre-existing cracks, with only a few exceptions.
The choice of refinement indicator is critical to minimizing discretization error while maximizing computational efficiency. Most refinement criteria proposed in the literature are largely heuristic. Heister et al. \cite{heister2015prisms}, for example, refined regions in which the phase field falls below a prescribed threshold. Criteria of this kind do not generalize to crack-nucleation problems, because nucleation is governed by the material strength and the phase field remains uniformly equal to one until fracture initiates. Energetic criteria have been introduced to overcome this limitation. Gupta et al.\ \cite{Gupta2022_AdaptiveMeshRefinement} proposed a multi-level mark–unmark strategy in which elements are first marked by the inter-iteration increment in strain-energy density, which locates the incipient crack, and subsequently by the phase field once damage has developed. Hirshikesh et al. \cite{hirshikesh2021adaptive} used linear stability analysis to identify the onset of rapid phase-field evolution. 
Nevertheless, there remains a need for a simple, physics-based, and broadly applicable refinement indicator that can be used consistently across both crack-nucleation and crack-propagation problems and for a wide range of material behaviors.

Even with AMR, the computational cost can continue to increase substantially for heterogeneous materials and large structures involving distributed crack nucleation and propagation, because the finely resolved phase-field region grows with the total crack surface area. This challenge becomes more severe for nonlinear material behavior and for problems such as fatigue, in which cracks advance incrementally over thousands of loading cycles. Ideally, the mesh would be fully coarsened in the crack wake so that the overall problem size remains approximately constant as the crack surface grows. Here, the crack wake refers not only to the surrounding region of crack that was previously refined but also the fractured regions behind the crack tip, where the phase field satisfies ${v}<1$.

Despite the extensive literature on AMR for phase-field fracture, adaptive coarsening has received comparatively little attention. Wick \cite{wick2016goal} and Patil et al. \cite{patil2018adaptive} considered mesh coarsening in regions surrounding the crack wake only where the phase field has its intact value, ${v}=1$. 
Several combined phase field-G/X-FEM approaches have been proposed \cite{giovanardi2017xfield, geelen2020extended, muixi2021combined}
which introduce Heaviside enrichments to represent the crack surface behind the crack tip. Although this approach can achieve the goal of keeping computational cost constant, it requires the implementation of enriched finite element technology, often with a global-local coupling framework, neither of which is straightforward to incorporate into many finite element libraries. Global-local methods also introduce additional modeling choices, including selecting an appropriate local-domain size and specifying suitable boundary conditions along the local-domain boundary.

In this context, the key contributions of this paper are:
\begin{enumerate}[leftmargin=*]
\item We develop an adaptive mesh-coarsening strategy in which the fractured region in the crack wake is replaced by a crack band that preserves the essential mechanical characteristics of a sharp crack in tension and shear and accurately reproduces the crack-opening displacement.
\item We introduce a physics-based adaptive mesh-refinement indicator derived from the material strength surface. Because violation of the strength surface is a necessary condition for crack growth, it provides a natural criterion for identifying regions in which fracture is imminent. The indicator applies uniformly to both crack nucleation and propagation across all brittle materials.
\item We present a parallel, scalable implementation in FEniCSx using PETSc solvers, \texttt{mpi4py} communication, and the parallel mesh-refinement capabilities of DOLFINx, thereby enabling large-scale simulations on distributed-memory computing platforms. The implementation is made publicly available as open-source software. 
To the best of our knowledge, existing open-source implementations are limited to serial execution and are based on legacy FEniCS \cite{freddi2023adaptive, Gupta2022_AdaptiveMeshRefinement}, which restricts their applicability to large-scale problems.
\item We assess the generality of the proposed adaptive refinement and coarsening framework through a broad set of benchmark problems, including brittle-fracture nucleation and propagation under quasi-static and dynamic loading, linear- and finite-elastic material behavior, and thermomechanical fracture.
\end{enumerate}

The remainder of this paper is organized as follows. Section~\ref{Sec:Theory} reviews the phase-field fracture theory. Section~\ref{Sec:AMR} describes the adaptive mesh refinement and coarsening algorithm. Section~\ref{Sec:Implementation} discusses the FEniCSx implementation. Section~\ref{Sec:Results} presents numerical results on the several benchmark problems and demonstrates the computational saving associated with the proposed adaptive approach. 
Section~\ref{Sec:Conclusions} provides a future outlook on combining adaptive mesh refinement and coarsening with acceleration of the staggered scheme and adaptive time stepping, along with some preliminary results and concluding remarks.

\section{The phase-field fracture theory}\label{Sec:Theory}

In this section, we briefly review the phase-field approach to fracture. We use the phase-field approach of Kumar et al.\ \cite{KFLP18, KBFLP20} because it has been demonstrated to be \emph{complete} through extensive validation in predicting fracture nucleation and propagation in elastic brittle materials under arbitrary monotonic, quasi-static loading. However, the adaptive refinement and coarsening strategy presented in Section \ref{Sec:AMR} applies to all other phase field approaches as long as a strength surface is well defined in that approach \cite{bourdin2025variational, vicentini2025variational}. This is commonly the case when the AT$_1$ regularization of surface energy is utilized.
Beyond quasi-static brittle fracture, neither the approach of Kumar et al. nor alternative formulations have been fully validated. We therefore adopt straightforward extensions that have shown promising results. However, again, the refinement and coarsening strategy does not critically depend on the specific formulation employed.

\subsection{Phase-field model for brittle materials under quasi-static loading}

Consider an elastic body occupying a domain $\Omega \subset \R^N$ ($N = 2$ or $3$) in its reference configuration. Under external loads,  the body experiences a deformation field $\bfy(\bfX,t)$ and displacement field $\bfu(\bfX,t)$.
Subject to the appropriate initial and boundary conditions, and in the absence of body forces and inertia, the deformation field $\bfy_k(\bfX)=\bfy(\bfX,t_k)$ and the phase field $v_k(\bfX)=v(\bfX,t_k)$ at any material point $\bf X\in\overline{\mathrm{\Omega}}$ and discrete time $t_k\in\{0=t_0,t_1,...,t_m,$ $t_{m+1},...,$ $t_M=T\}$ are obtained from solving system of coupled partial differential equations (PDEs)
{
\begin{equation}
\left\{\begin{array}{lll}
\hspace{-0.15cm} {\rm Div} \left[v_{k}^2  \dfrac{\partial {W}}{\partial \bfF}(\nabla \bfy_k) \right]=0 ,
\vspace{0.1cm}\\
\hspace{-0.15cm}
\dfrac{3}{4} \varepsilon \, \de \,  G_c \triangle v_{k}=2 v_{k}\, {W}(\nabla\bfy_k)  +c_\texttt{e} -\dfrac{3}{8} \dfrac{ {\delta^\varepsilon}  \, G_c }{\varepsilon}, & \text{if } 0<v_k<\overline{v}_k,
\vspace{0.1cm}
\end{array}\right. \label{phase-field-equations}
\end{equation}
where $W$ is the strain energy density function, $\bfF=\nabla \bfy$ is the deformation gradient tensor, $\varepsilon$ is a regularization length, and $G_c$ is the fracture toughness. The second relation in (\ref{phase-field-equations}) states the evolution equation for the phase field as a bound-constrained problem. 
The phase field is restricted to the admissible range $0\le v_k\le\overline{v}_k$,
where the upper bound
\begin{equation}
\overline{v}_k(\bfX)=\left\{\begin{array}{ll} v_{k-1}(\bfX), & \text{if } v_{k-1}(\bfX)\le v_a,\\[4pt] 1, & \text{otherwise},\end{array}\right.
\label{eq:bounds}
\end{equation}
enforces the irreversibility of fracture: at material points where the phase field has dropped below the threshold $v_a=0.05$, it is not allowed to increase again, so that a fully developed crack cannot heal. Implementing weak irreversibility in this manner allows the crack to fully develop and yield the most physical results \citep{dolbow2025uniform}.

}

The constitutive prescription for the driving force $c_\texttt{e}(\bfX,t)$ in (\ref{phase-field-equations})$_2$ depends on the strength surface of the material. The formulation allows for an arbitrary choice of the strength surface as shown recently by Chockalingam et al. \cite{chockalingam2025MCHB}. In this work, we have adopted the Drucker-Prager (DP) surface, expressed as follows:
\begin{equation}
	\mathcal{F}(\bfS)=\sqrt{J_2}+\gamma_1 I_1+\gamma_0=0 ,\label{DP}
\end{equation}
with
\begin{equation}
	\gamma_0=-\sss, \qquad  \gamma_1=\dfrac{\sqrt{3} \sss-\sts}
{\sqrt{3}\sts},
\end{equation}
where
\begin{equation}
	I_1 = \mathrm{tr}\,\bfS, \quad
J_2 = \frac{1}{2}\mathrm{tr}\,\bfS_D^2, \quad
\bfS_D = \bfS - \frac{1}{3} (\mathrm{tr}\,\bfS) \bfI ,\label{T-invariants}
\end{equation}
stand for two of the standard invariants of the stress tensor $\bfS$, while the constants $\sts>0$ and $\sss>0$ denote the uniaxial tensile and shear strengths of the material, respectively. Under finite strains, we define the strength surface in terms of the Biot stress tensor $\bfS=({\bfS^{(1)}}^T\bfR+\bfR^T\bfS^{(1)})/2$, where $\bfR$ is the rigid rotation tensor defined through a polar decomposition of the deformation gradient $\bfF=\bfR\bfU$, with $\bfU$ being the right stretch tensor \cite{KKLP24} and
\begin{equation}
{\bfS}^{(1)}(\bfX,t)=\dfrac{\partial {W}}{\partial \bfF}(\bfF).
\end{equation}
is the first Piola-Kirchhoff stress tensor. In linear elasticity, the strength surface is defined in terms of the linearized stress tensor
\begin{equation*}
\boldsymbol{\sigma}(\bfX,t)=\dfrac{\partial W}{\partial \bfE}(\bfE(\bfu))
\end{equation*}
where the linearized strain is $\bfE = \text{sym}(\nabla \mathbf{u})$.

Corresponding to the DP surface, the functional form for $c_\texttt{e}(\bfX,t)$ was presented in \cite{KRLP22}:
\begin{align}
c_{\texttt{e}}(\bfX,t)=\beta_2^\varepsilon\sqrt{\mathcal{J}_2}+\beta_1^\varepsilon \mathcal{I}_1-
v (1-sgn(\mathcal{I}_1)) \, W(\bfF),\label{cehat-2022}
\end{align}
with
\begin{equation}
\left\{\begin{array}{l}
\beta^\varepsilon_1=\dfrac{1}{\shs}\delta^\varepsilon\dfrac{G_c}{8\varepsilon}-\dfrac{2\mathcal{W}_{\texttt{hs}}}{3\shs}\vspace{0.2cm}\\
\beta^\varepsilon_2=\dfrac{\sqrt{3}(3\shs-\sts)}{\shs\sts}\delta^\varepsilon\dfrac{G_c}{8\varepsilon}+
\dfrac{2\mathcal{W}_{\texttt{hs}}}{\sqrt{3}\shs}-\dfrac{2\sqrt{3}\mathcal{W}_{\texttt{ts}}}{\sts}\end{array}\right.  \label{betas}
\end{equation}
Here, ${W}_{\texttt{ts}}$ and ${W}_{\texttt{hs}}$ stand for the values of the strain energy function along uniform uniaxial tension and hydrostatic stress states at which the strength surface is violated. $\shs$ is the hydrostatic strength which for the DP surface is related to shear and tensile strengths through the relation
\begin{equation}
    \shs= \dfrac{\sss \sts}{3 \sss - \sqrt{3} \sts}.
    \label{shs-sts-sss}
\end{equation}
$\mathcal{I}_1$ and $\mathcal{J}_2$ stand for the invariants of the degraded Biot stress.
The coefficient $\delta^{\varepsilon}$ is given by
\begin{equation}
\delta^{\varepsilon} = \left(1 + \frac{3h}{8\varepsilon}\right)^{-2} \left(\frac{\sts + (1 + 2\sqrt{3})\shs}{(8 + 3\sqrt{3})\shs}\right) \frac{3G_c}{16\mathcal{W}_{\texttt{ts}}\varepsilon} + \left(1 + \frac{3h}{8\varepsilon}\right)^{-1} \frac{2}{5},
\label{eq:delta}
\end{equation}
where $h$ denotes the finite element size.

The constitutive prescription for the nonlinear strain energy function is made as follows per the prescription from \cite{LP10}:
\begin{equation}
\begin{aligned}
    {W}(\bfF)=&\sum_{r=1}^{2} \frac{3^{1-\alpha_r}}{2\alpha_r} \mu_r \left[ (\mathbf{F} \cdot \mathbf{F})^{\alpha_r} - 3^{\alpha_r} \right]
    - \sum_{r=1}^{2} \mu_r \ln (\det \mathbf{F}) + \frac{\kappa}{2} (\det \mathbf{F} - 1)^2,
\end{aligned}
\label{eq:energy_function}
\end{equation}
where, $\mu_1$ and $\mu_2$ are shear modulus parameters, such as total shear modulus $\mu= \mu_1 +\mu_2$, $\kappa$ is the bulk modulus, and $\alpha_1$, $\alpha_2$ are strain stiffening parameters. In linear, isotropic elasticity, this reduces to
\begin{equation}
	{W}(\bfE(\bfu)) =\frac{E}{2(1+\nu)} \, {\rm tr}\,\bfE^2+\dfrac{E\nu}{2 (1+\nu)(1-2\nu)}({\rm tr}\,\bfE)^2,\label{W-mu}
\end{equation}
where $E$ is the Young's modulus and $\nu$ is the Poisson's ratio.

\subsubsection{The phase-field strength surface indicator}\label{Sec:Indicator}

The governing equations (\ref{phase-field-equations}) are constructed by modification of the AT$_1$ classical variational model. As shown in \cite{KBFLP20}, these equations dictate that phase field can only evolve from its initial value of $v=1$ when
\begin{equation}
\mathcal{I}(\mathbf{X}, t) = 2 \, W(\bfF) + \ce - \dfrac{3 \de G_c}{8 \eps}=0 \label{strength-surface-ce}
\end{equation}
obtained by setting $v=1$ in the right-hand side of the evolution equation for phase-field (\ref{phase-field-equations})$_2$. {The surface $\mathcal{I}(\mathbf{X}, t)=0$} can be regarded as a phase-field approximation of the strength surface ${\mathcal{F}}(\boldsymbol{\sigma})$ (\ref{DP}) and it signals the onset of fracture nucleation.
The phase-field strength surface reduces to the exact strength surface for $\eps \searrow 0$ \cite{KBFLP20}. Note that setting $\ce=0$ and $\de=1$ yields the phase field strength surface for the classical variational AT$_1$ model \cite{pham2011gradient}.

Since a crack cannot nucleate at a point unless $\mathcal{I}(\mathbf{X}, t) =0$, it serves as a natural threshold indicator for adaptive mesh refinement, as we will discuss in Section \ref{Sec:AMR}.

\subsection{Phase-field model for dynamic fracture in brittle materials}

As discussed in a recent work \cite{dahal2026dynamic}, a simple extension of the quasi-static brittle fracture model \eqref{phase-field-equations} is add the inertial term to the balance of linear momentum. While this formulation may still miss some essential physics such as the crack-velocity-dependence of material properties, it agrees well with a host of benchmark problems such as the Kalthoff-Winkler test and the dynamic branching test; also see Liu et al. \cite{liu2024effects}. The modified formulation stated in linear elasticity for which it has previously been evaluated is as follows:
subject to the appropriate initial and boundary conditions, the displacement field $\mathbf{u}_k(\mathbf{X})=\mathbf{u}(\mathbf{X},t_k)$ and phase-field $v_k(\mathbf{X})=v(\mathbf{X},t_k)$ at any material point $\mathbf{X}\in\overline{\Omega}$ and discrete time $t_k\in\{0=t_0,t_1,\dots,t_m,$ $t_{m+1},\dots,$ $t_M=T\}$ are determined by the system of coupled partial differential equations
\begin{equation}
\left\{\begin{array}{ll}
{\rm div}\left[v_{k}^2 \dfrac{\partial W}{\partial \bfE}(\bfE(\bfu_{k}))\right]
= v_{k}^{b} \,\rho\,\ddot{\mathbf{u}}_{k},\\[10pt]
\dfrac{3}{4} \varepsilon \, \de \,  G_c \triangle v_{k}=2 v_{k} W(\bfE(\bfu_{k}))+c_\texttt{e}(\bfX,t_{k})- \dfrac{3}{8}  \dfrac{\de \, G_c}{\varepsilon}, \\[10pt]
\end{array}\right.
\label{phase-field-dynamics}
\end{equation}
{where the phase-field equation is  subject to the bounds (\ref{eq:bounds}), and} $\ddot{\mathbf{u}} = {\partial^2 \mathbf{u}}/{\partial t^2}$ stands for the acceleration field. The exponent $b$ controls the degradation of the inertial term. Most dynamic fracture formulations in the literature \cite{borden2012phase, Hofacker2012_ContinuumPhaseField, Nguyen2018_PhaseFieldCohesive, Geelen2019_PhaseField} have assumed $b=0$ based on the argument that degrading the density would violate the mass conservation law. Dahal et al. \cite{dahal2026dynamic} have recently argued for $b=2$ based on the argument that the density needs to be degraded for the fractured region to not get artificially distorted and widen because of the presence of kinetic energy. Since in the approach of Kumar et al., the regularization length $\varepsilon$ is a free parameter, they argued that degrading density only introduces a regularization error in the conservation of mass on par with the error introduced in elastic behavior by smearing a sharp crack over a finite length scale. In the numerical example presented in Section \ref{Sec:Branching}, we adopt $b=2$. The density degradation formulation for the classical variational model has also been recently evaluated by Durussel et al. \cite{durussel2026dynamic}, del Castillo and Li \cite{del2026dynamic} and Heinzmann et al. \cite{Lorenzis2026dynamic}. The phase field strength surface indicator remains the same as in the quasi-static formulation. The adaptive formulation is independent of the choice of $b$.

\subsection{Phase-field model for thermo-mechanical fracture in brittle materials}

Following Zeng and Dolbow \cite{zengdolbow2026thermal}, the governing equations for thermo-mechanical fracture in brittle materials under quasi-static loading are constructed by simply coupling the thermal formulation with the mechanical formulation described in the previous subsection. The resulting formulation was shown to be capable of accurately predicting brittle fracture in materials subjected to severe thermal shock.

The temperature field is denoted by $T(\bfX, t)$. $c_p$ denotes the specific heat, $\boldsymbol{\kappa}$ the thermal conductivity tensor, and $s$ is the volumetric heat source. The infinitesimal strain tensor is decomposed into a mechanical strain and thermal strain:
\begin{equation}
    \bfE = \bfE_{m} + \bfE_T
\end{equation}
with the thermal strain constitutively prescribed as
$$\mathbf{E}_T = \alpha \left( T(\mathbf{X}, t) - \tilde{T} \right) \mathbf{I}$$
where $\tilde{T}$ is the reference temperature and $\alpha$ is the coefficient of thermal expansion. The linearized stress tensor is defined in terms of the mechanical stored energy $W_m$ as
\begin{equation*}
\boldsymbol{\sigma}(\bfX,t)=\dfrac{\partial W_m}{\partial \bfE_m}(\bfE_m(\bfu, T)).
\end{equation*}

The displacement field $\bfu_k(\bfX)=\bfu(\bfX,t_k)$, temperature field $T_k(\bfX)=T(\bfX,t_k)$  and the phase field $v_k(\bfX)=v(\bfX,t_k)$ at any material point $\bf X\in\overline{\mathrm{\Omega}}$ and discrete time $t_k\in\{0=t_0,t_1,...,t_m,$ $t_{m+1},...,$ $t_M=T\}$ are obtained from solving system of coupled partial differential equations (PDEs)
\begin{equation}
\left\{\begin{array}{ll}
\hspace{-0.15cm} {\rm div} \left[v_{k}^2  \dfrac{\partial W_m}{\partial \bfE_m}(\bfE_m(\bfu_{k}, T_k))\right]=0 ,
\vspace{0.1cm}\\
\hspace{-0.15cm}
\rho c_p \dfrac{\partial T_k}{\partial t} = {\rm div} (\boldsymbol{\kappa} \cdot \nabla T_k) + s ,
\vspace{0.1cm}\\
\hspace{-0.15cm}
\dfrac{3}{4} \varepsilon \, \de \,  G_c \triangle v_{k}=2 v_{k}\, {W_m}(\bfE_m(\bfu_{k}, T_k))  +c_\texttt{e} -\dfrac{3}{8} \dfrac{ {\delta^\varepsilon}  \, G_c }{\varepsilon},
\end{array}\right. \label{phase-field-thermal}
\end{equation}
subject to the appropriate Dirichlet and Neumann boundary conditions on the displacement, temperature, and phase fields{, and with the phase-field equation again subject to the bounds (\ref{eq:bounds})}. The driving force $c_\texttt{e}$ and coefficient $\de$ remain the same as for the purely mechanical problem.
A more detailed description of the formulation is provided by Zeng and Dolbow \cite{zengdolbow2026thermal}.

\section{Adaptive mesh refinement and coarsening}\label{Sec:AMR}

In this section, we present a novel strategy for adaptive spatial discretization of the governing equations of phase-field fracture. An indicator identifies, before the phase field starts to evolve, the material points where the strength surface is about to be reached. A small zone around these points is refined from the base coarse mesh to a fine mesh. Later, once a phase-field crack fully develops in that region, and the front has moved on, the indicator separates the crack tip from the wake of the crack so that the wake can return to the original coarse mesh size. The fine discretization therefore remains localized around the crack tip.

In Section \ref{Sec:BaseMesh}, we first introduce the meshes involved, then the refinement and coarsening indicators (Section \ref{Sec:IndicatorAMR}), the construction of the refined mesh (Section \ref{Sec:Refinement}), and the coarsening of the wake (Section \ref{Sec:Coarsening}). The complete procedure is summarized in Algorithms \ref{alg:step} and \ref{alg:remesh} and is illustrated in Fig. \ref{Fig0}.

\subsection{Meshes and refined window}\label{Sec:BaseMesh}

\begin{figure}[h]
    \centering
    \includegraphics[width=6in]{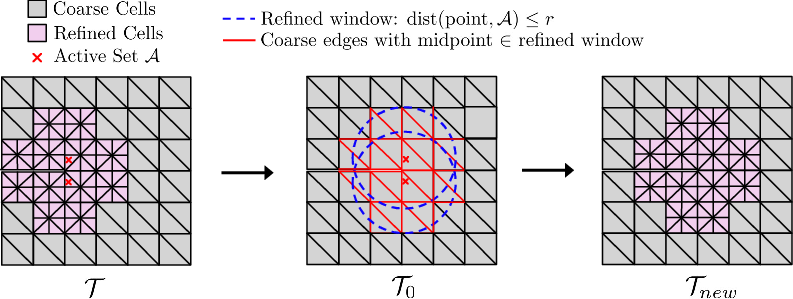}
    \caption{Schematic of the adaptive mesh refinement and coarsening algorithm. The algorithm finds the active set on  $\Tcal$ and refines coarse edges with midpoints within the refined window on $\Tcal_0$. More details of the algorithm are in Section 3.3.}\label{Fig0}
\end{figure}

The adaptive spatial discretization algorithm utilizes three meshes:

\begin{itemize}[leftmargin=*]

\item the \emph{base mesh} $\Tcal_0$: the fixed coarse mesh of size $h_0$ 
of simplicial elements (triangles or tetrahedra). It is assumed that the base mesh size
is sufficiently small for an accurate elastic response. The mesh size in the crack band obtained
after coarsening is also assumed to be equal to the base mesh size throughout this work for simplicity. Therefore, the base mesh size is chosen to be of the order of the regularization length and small enough relative to the characteristic structural length scale.
Wherever the phase field is about to evolve, the base mesh is refined to a \emph{target size} $\hmin=h_0/2^{n}$ by $n$ levels of edge bisection. For instance, $n=2$ gives $\hmin=\varepsilon/4$ for $h_0=\varepsilon$;

\item the \emph{current mesh} $\Tcal$: the mesh on which the displacement, the phase field, and all variables currently live and on which the governing equations are being solved. It is the mesh produced by the most recent remeshing event;

\item the \emph{new mesh} $\Tcal_{\rm new}$: the mesh built by the next remeshing event. It is always constructed from the base mesh $\Tcal_0$ by refining $\Tcal_0$ around the centroids of cells that the refinement indicator presented in the next subsection currently flags on $\Tcal$, as shown in Fig. \ref{Fig0}. Once the fields have been transferred from $\Tcal$ to $\Tcal_{\rm new}$, the new mesh becomes the current mesh, $\Tcal\gets\Tcal_{\rm new}$, and the old current mesh is discarded.

\end{itemize}

Building $\Tcal_{\rm new}$ from $\Tcal_0$ rather than from $\Tcal$ is what makes the coarsening possible, as discussed in Section \ref{Sec:Coarsening}. Because every mesh is obtained from $\Tcal_0$ by edge bisection, the vertices of $\Tcal_0$ are vertices of $\Tcal$ and of $\Tcal_{\rm new}$, which is the property that the field transfer of Section \ref{Sec:Coarsening} exploits.

The refined region, which we call the \emph{refined window}, is the set of points within a distance $r$ of the centroids of cells flagged by the indicator presented in the next subsection, as illustrated in Fig. \ref{Fig0}.
We use $r=3 \, \varepsilon$ throughout. A window of this size contains the entire regularized crack profile around the current front for the chosen AT$_1$ regularization and leaves a margin of about one regularization length ahead of the flagged cells, so that the front can advance during several load steps before the mesh has to be modified. Larger values of $r$ reduce the number of remeshing events at the expense of a larger window.

\subsection{The strength-surface refinement and coarsening indicator}\label{Sec:IndicatorAMR}

As discussed in Section \ref{Sec:Indicator}, phase field evolution from $v=1$ can only occur when the phase field strength surface $\mathcal{I}(\mathbf{X}, t)$ is reached. We make use of this fact to derive the following dimensionless indicator for mesh refinement:
{
\begin{equation}
\Ibar(\bfX,t)=\dfrac{\mathcal{I}(\bfX,t)}{\dfrac{3}{8}\dfrac{\de G_c}{\varepsilon}}=\dfrac{2\,W+\ce\big|_{v=1}}{\dfrac{3}{8}\dfrac{\de G_c}{\varepsilon}}-1,
\label{eq:indicator}
\end{equation}
evaluated with the current displacement field and with the phase field set to one, that is, with the undegraded stress invariants in the driving force $\ce$ of (\ref{cehat-2022}). The indicator can be used with the classical AT$_1$ variational model by setting $\ce=0$ and $\de=1$.  $\Ibar=-1$ corresponds to the unstressed state, $\Ibar\geq0$ corresponds to the violation of strength surface. 
A cell $K$ of the current mesh is flagged as \emph{active} when
\begin{equation}
\Ibar(\bfX_K,t)>-\theta,
\label{eq:active}
\end{equation}
where $\bfX_K$ is the centroid of the cell and $\theta$ is a safety margin which we set to $\theta=0.1$ in all computations. In other words, a cell is marked for refinement as soon as the driving force at its centroid reaches 90\% of the strength resistance. {We denote by $\Acal$ the set of centroids of active cells of $\Tcal$. The refined window of Section \ref{Sec:BaseMesh} is the set of points of the body within the distance $r=3\varepsilon$ of $\Acal$, and the mesh $\Tcal_0$ is refined by bisecting every edge whose midpoint lies within the window and whose length exceeds the target size; more details are in Section \ref{Sec:Refinement}.} Figure \ref{Fig1}(a) illustrates the  the active cells and the refined window for a crack propagating in the single-edge notched plate, studied in detail later in Section \ref{Sec:SENT}. The active cells fill the phase field crack along its entire length and the mesh remains refined all along it as it propagates.

The advantage of defining the refinement indicator based on strength surface rather than other heuristic indicators in past work is immediately apparent. It is a simple algebraic expression that enters the equation and its use does not require additional computational work. Violation of strength is a necessary condition for crack evolution in every setting, so it is broadly applicable across all nucleation and propagation problems for arbitrary linear or finite elastic material and for any quasi-static or dynamic loading. Furthermore, unlike criterion based on strain energy density \cite{Gupta2022_AdaptiveMeshRefinement, freddi2022mesh} that only predict \emph{where} the crack growth is most likely to occur, the indicator based on strength naturally provides a measure of both \emph{where} and \emph{when} it will occur. Strength naturally provides a tension-compression asymmetry \cite{LK24} that prevents false flagging and refinement in regions of large compression where failure in brittle materials seldom happens.

{
Coarsening of the mesh in the crack wake requires an indicator function to separate that region from the crack tip where a fine mesh is needed to correctly drive crack growth.
We make use of the insight that the flanks of a fully developed crack behind the crack tip, where $v_{\texttt{c}}\le v<1$, are unloaded because the crack faces are traction-free, and they are not active anymore according to (\ref{eq:active}). So if we can additionally make inactive the cells in the core of a phase field crack, where $v<v_{\texttt{c}}$, then only active cells will be in front of the crack tip.
Therefore, we modify the indicator in (\ref{eq:indicator}) as
\begin{equation}
\Ibar_{\texttt{c}}(\bfX,t)=\mathcal{H}(v-v_{\texttt{c}})\,\dfrac{2\,W+\ce\big|_{v=1}}{\dfrac{3}{8}\dfrac{\de G_c}{\varepsilon}}-1,\qquad v_{\texttt{c}}=0.6,
\label{eq:indicator-coarsening}
\end{equation}
with
\begin{equation}
\mathcal{H}(x)=\left\{\begin{array}{ll} 1, & \text{if } x\ge 0,\\[4pt] 0, & \text{if } x<0,\end{array}\right.
\label{eq:heaviside}
\end{equation}
so that $\Ibar_{\texttt{c}}=-1$ in the core of a fully developed crack. Fig.\ \ref{Fig1}(b) shows that with this indicator, the active cells form a thin crescent ahead of the tip. The refined window is the disk of radius $3\varepsilon$ around them, and the wake can be returned to the base mesh.

}

\begin{figure}[htbp]
    \centering
    \includegraphics[width=5.5in]{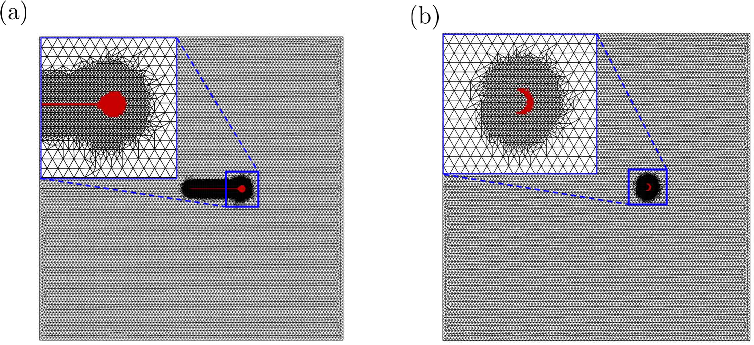}
    \caption{{Refined window and active cells during crack propagation in the single-edge notched plate of Section \ref{Sec:SENT}, (a) without coarsening (AMR) and (b) with coarsening (AMRC). The dark regions are the refined parts of the mesh; the insets show the active cells (red), which the refined window surrounds to a distance of $3\varepsilon$.}}\label{Fig1}
\end{figure}

\subsection{Construction of the refined mesh}\label{Sec:Refinement}

The indicator (\ref{eq:indicator}) or (\ref{eq:indicator-coarsening}) is evaluated based on the fields in the current mesh $\Tcal$ to obtain the set $\Acal$ of centroids of the active cells that satisfy (\ref{eq:active}). Once computed, the active set of centroidal positions is carried over to the base mesh $\Tcal_0$ as purely geometric information.
Given this set $\Acal$, the refined mesh is constructed from the base mesh by $n$ successive passes of edge bisection, as described in Algorithm \ref{alg:remesh}. In pass $\ell$, every edge of the intermediate mesh produced by pass $\ell-1$, starting from $\Tcal_0$ in the first pass, whose midpoint lies within the distance $r$ of $\Acal$ and whose length exceeds $1.05\,h_0/2^{\ell}$ is marked, and the marked edges are bisected following the algorithm of Plaza and Carey \cite{plaza2000mesh}. The algorithm bisects the marked edges and then bisects as many additional edges as needed to remove hanging nodes, producing a conforming mesh. 
Outside the refinement window, the closure of the bisection produces a graded transition back to $h_0$ over a layer of a few elements. The distance queries are performed with a k-d tree of the active centroids \cite{manewaldscipy2020}, so that the marking costs a logarithmic factor per edge. The mesh delivered by the last pass is $\Tcal_{\rm new}$.

Every mesh generated in this way is nested in the base mesh. The motivation for always refining from the base mesh is to make the mesh coarsening possible in the FEniCSx implementation, as discussed in the following subsection. This adds additional cost compared to refining from the constructed mesh in the last step; however, in our computations, the additional cost is small relative to the cost of solving the phase field equations once.

Refinement is triggered whenever at least one active cell has a diameter larger than $1.5\,\hmin$, that is, when the active set reaches the coarse part of the mesh. This happens when a crack front approaches the edge of its window, or when a new region of the body approaches the strength surface. The test is performed inside the staggered loop, after every solution of the displacement problem and before the solution of the phase-field problem (Algorithm \ref{alg:step}). This ensures that the phase field is never solved on a cell that is active but coarse, so that its evolution, whether by nucleation or by propagation, always takes place at the resolution $\hmin$. Combined with the margin $\theta$, it also ensures that refinement precedes nucleation. A moderate window size of $r= 3 \, \varepsilon$ ensures that this remeshing does not occur too frequently, especially for stably propagating cracks. After a remeshing, the staggered iteration simply continues on the new mesh $\Tcal_{\rm new}$, which becomes the current mesh $\Tcal$, with the transferred fields as its starting point.

\begin{algorithm}[h]
\caption{Adaptive staggered solution of one load or time step $t_k$.}\label{alg:step}
\begin{algorithmic}[1]
\Require base mesh $\Tcal_0$, current mesh $\Tcal$, fields $(\bfu_{k-1},v_{k-1})$ on $\Tcal$
\State Apply the boundary condition of $t_k$; set the lower bound $v_{\rm lb}=0$ and the upper bound $v_{\rm ub}=\overline{v}_k$ of (\ref{eq:bounds})
\State $i\gets 0$, $v^{(0)}\gets v_{k-1}$
\Repeat
  \State $i\gets i+1$
  \State Solve the displacement problem (\ref{phase-field-equations})$_1$ on $\Tcal$ for $\bfu^{(i)}$ with $v^{(i-1)}$ fixed
  \State Evaluate $\Ibar$ at the centroid of every cell of $\Tcal$ and form the active set $\Acal$ \Comment{Eqs.\ (\ref{eq:indicator})--(\ref{eq:indicator-coarsening})}
    \If{some active cell of $\Tcal$ has diameter $h_K>1.5\,\hmin$}
       \State $\Tcal_{\rm new}\gets\textsc{Remesh}(\Tcal_0,\Acal)$ \Comment{Algorithm \ref{alg:remesh}}
     \State Interpolate $\bfu^{(i)}$, $v^{(i-1)}$, the bounds $v_{\rm lb}$, $v_{\rm ub}$, and the history variables from $\Tcal$ onto $\Tcal_{\rm new}$
     \State Rebuild the forms and solvers on $\Tcal_{\rm new}$ and reapply the boundary condition of $t_k$
     \State $\Tcal\gets\Tcal_{\rm new}$
  \EndIf
  \State Solve the phase-field problem (\ref{phase-field-equations})$_2$ on $\Tcal$ for $v^{(i)}$ with $\bfu^{(i)}$ fixed
\Until{ {$\|v^{(i)}-v^{(i-1)}\|_{L^2(\Omega)}<\texttt{tol}$ or $i=i_{\max}$} }
\State $(\bfu_k,v_k)\gets(\bfu^{(i)},v^{(i)})$; post-process and write output
\end{algorithmic}
\end{algorithm}

\begin{algorithm}[h]
\caption{\textsc{Remesh}: construction of the new mesh $\Tcal_{\rm new}$ from the base mesh $\Tcal_0$.}\label{alg:remesh}
\begin{algorithmic}[1]
\Require base mesh $\Tcal_0$ of size $h_0$, active centroids $\Acal$ of the cells of the current mesh $\Tcal$ flagged by (\ref{eq:active}), gathered from all processes, radius $r$, target size $\hmin$
\State $\Tcal_{\rm new}\gets\Tcal_0$; $n\gets\lceil\log_2(h_0/\hmin)\rceil$; build a k-d tree of $\Acal$
\For{$\ell=1,\dots,n$}
   \State $\mathcal{E}\gets\{\,e\in\text{edges}(\Tcal_{\rm new}):\ |e|>1.05\,h_0/2^{\ell}\ \text{and}\ \operatorname{dist}(\text{midpoint}(e),\Acal)\le r\,\}$
   \If{$\mathcal{E}=\emptyset$ on all processes} \textbf{break} \EndIf
   \State $\Tcal_{\rm new}\gets\textsc{Bisect}(\Tcal_{\rm new},\mathcal{E})$
   \Comment{conforming Plaza refinement, in parallel}
\EndFor
\Ensure $\Tcal_{\rm new}$
\end{algorithmic}
\end{algorithm}

\subsection{Construction of the coarsened mesh}\label{Sec:Coarsening}

The mesh coarsening follows a simple strategy. The indicator (\ref{eq:indicator-coarsening}) masks the cells of the crack wake from the active set. The new mesh $\Tcal_{\rm new}$ is built from the base mesh $\Tcal_0$ by refining only around the cells that are active at the crack tip, so a region that was refined at an earlier remeshing event, and that the indicator no longer flags, is simply not refined this time and therefore returns to the base size $h_0$. Again, we are assuming the target mesh size in the crack wake equals the base mesh size for simplicity.
Moreover, as described in Section \ref{Sec:Refinement}, the marking is not transferred as a set of cells. The active set is reduced to the point cloud $\Acal$ of the centroids of the flagged cells of $\Tcal$ that no longer refers to any mesh. The edges of $\Tcal_0$ are then marked by the distance query $\operatorname{dist}(\text{midpoint}(e),\Acal)\le r$ of Algorithm \ref{alg:remesh}. No mapping between the cells of $\Tcal$ and the cells of $\Tcal_0$ is needed.
This differs from the usual practice of coarsening by de-refining parent cells, which requires storing and traversing the refinement history \cite{kirk2006libmesh, arndt2021dealii, kim2024octree} and is not available in FEniCSx or many other libraries\footnote{In finite element libraries such as MFEM or MOOSE where coarsening option is directly available, one could utilize either the default approach or our approach with the rest of the algorithm remaining the same.}. It also differs from strategies that only coarsen the intact material surrounding the crack \cite{wick2016goal, patil2018adaptive}; here, the fractured material itself is coarsened.

The fields are then transferred from the current mesh $\Tcal$ onto $\Tcal_{\rm new}$. Since the vertices of the base mesh are vertices of every adaptive mesh, the transfer of fields (displacement, phase field, velocity, acceleration, temperature) to a coarsened region amounts to restricting the nodal values to the base-mesh nodes, whereas in a newly refined region it amounts to interpolating the old piecewise-linear fields.
Spatially varying material properties, such as the random strength fields of Sections \ref{Sec:Uniaxial} and \ref{Sec:Thermal}, are re-sampled from their definition on the base mesh or as a function of position, so that they are identical on every mesh.

After coarsening, the phase-field crack no longer has the optimal profile corresponding to AT$_1$ regularization \cite{KBFLP20}. The phase-field at the coarse element nodes takes the values of phase-field in the core of the fully developed crack i.e., close to 0 in one layer of elements. The irreversibility constraint ensures that these values remain pinned. The result is a band of fully degraded elements of width comparable to $h_0$, visible in the inset of Fig.\ \ref{Fig2}(d), which we refer to as the \emph{crack band}. The band transmits neither normal nor shear tractions, and it opens freely. We will evaluate in Section \ref{Sec:Results} whether the resulting crack opening displacement equals the value for a refined mesh. Section \ref{Sec:Results} also evaluates whether coarsening preserves the energy response and what the maximum mesh size of the crack band is reasonable.

The crack band representation is reminiscent of the crack band approach of Ba\v{z}ant and Oh \cite{bazantoh1983}. However, in our approach, the phase-field equations resolve the crack physics on a fine mesh near the crack tip, and we use the crack band only in the wake. Therefore, the band width need not be tied to $G_c$ and is a discretization artifact with no significant mechanical consequence as shown in the results in Section 5. We also note that we do not currently account for the correct crack-face boundary conditions under compression. The directional split of the stress tensor proposed by Strobl and Seelig \cite{stroblseelig2015} and Steinke and Kaliske \cite{steinke2019} can be utilized for this purpose, applied only to the crack band in the wake of the crack in the equilibrium equation.

\section{Implementation in FEniCSx}\label{Sec:Implementation}

\subsection{Discretization and solvers}

All computations are performed with the open-source finite element library FEniCSx \cite{barrata2023dolfinx, logg2012automated}, specifically DOLFINx version 0.9, in which the weak forms of (\ref{phase-field-equations}), (\ref{phase-field-dynamics}), and (\ref{phase-field-thermal}) are written in the Unified Form Language and compiled just in time. The displacement and the phase field are approximated with continuous piecewise-linear Lagrange elements on simplices. In the finite-elasticity example of Section \ref{Sec:Slant}, the near incompressibility of the material calls for a locking-free discretization of the displacement; we use the nonconforming Crouzeix-Raviart element \cite{CrouzeixRaviart73}, stabilized by an interior-penalty term on the displacement jumps across element facets \cite{KFLP18}.

The coupled problem is solved at each load or time step by the staggered scheme of Algorithm \ref{alg:step} until the change of the phase field between two consecutive iterations, measured in the $L^2$ norm, falls below a prescribed tolerance. 
The phase-field problem is the bound-constrained problem (\ref{phase-field-equations}). The lower bound $v_{\rm lb}=0$ and the upper bound $v_{\rm ub}=\overline{v}_k$ of (\ref{eq:bounds}) are imposed on the nodal values at the beginning of every step, and the constrained problem is solved with the reduced-space active-set Newton method of the PETSc library \cite{balay2019petsc, benson2006flexible}, which treats the bounds exactly.

The linear systems are solved with the parallel direct solver MUMPS \cite{amestoy2001mumps} in the two-dimensional problems and in the three-dimensional echelon-crack problem. In the three-dimensional problem of Section \ref{Sec:Uniaxial}, where the refined mesh momentarily reaches about $1.7\times10^{6}$ tetrahedral elements, the displacement problem is solved with the conjugate gradient method preconditioned by the smoothed-aggregation algebraic multigrid of PETSc, while the scalar phase-field problem remains on the direct solver. In the finite-elasticity problem, {the displacement problem is solved with the conjugate gradient method preconditioned by BoomerAMG \cite{henson2002boomeramg}, and the phase-field problem with the active-set method and MUMPS}; when Newton's method fails to converge, which occurs at the onset of unstable crack growth in soft materials, the implicit gradient flow method is utilized \cite{KFLP18}.

In the dynamic problem, the momentum balance (\ref{phase-field-dynamics})$_1$ is integrated in time with the HHT-$\alpha$ method \cite{Hilber1977_ImprovedDissipation} with $\alpha_f=0.1$ and the corresponding Newmark parameters $\gamma=1/2+\alpha_f$ and $\beta=(1+\alpha_f)^2/4$.
In the thermomechanical problem, the heat equation (\ref{phase-field-thermal})$_2$ is integrated with a semi-implicit backward Euler scheme in which the temperature-dependent conductivity and heat capacity are evaluated at the temperature of the previous step and the convective boundary condition on the quenched edges imposed as a Robin condition.

\subsection{Adaptive discretization implementation}

Each bisection pass is a call to the parallel refinement routine of DOLFINx with the list of edges to be bisected. The indicators (\ref{eq:indicator}) and (\ref{eq:indicator-coarsening}) that mark active cells are UFL expressions of the current fields that are interpolated into a piecewise-constant space, which evaluates them at the cell centroids. The edge lengths and midpoints are computed from the mesh geometry with vectorized array operations, and the distance queries use the k-d tree of SciPy \cite{manewaldscipy2020}.

Field transfer uses DOLFINx's non-matching interpolation, which locates the nodes of $\Tcal_{\rm new}$ in the cells of $\Tcal$ via bounding-box trees and evaluates the fields of $\Tcal$ there. The problem definition (function spaces, boundary conditions, variational forms, and solvers) is built from a single function of the mesh. Remeshing therefore consists of calling this function on the new mesh, transferring the fields, and reusing the compiled kernels. The old mesh, function spaces, and solver objects are not saved after the transfer, so that the memory footprint is that of the current mesh plus the base mesh.

\section{Numerical results}\label{Sec:Results}

We now assess the framework on seven problems that exercise different aspects of fracture: propagation from a pre-existing crack under quasi-static mode-I loading (Sections \ref{Sec:SENT} and \ref{Sec:Surfing}), three-dimensional segmentation of a crack front (Section \ref{Sec:Echelon}), propagation in a nearly incompressible material under finite deformations (Section \ref{Sec:Slant}), dynamic propagation with branching (Section \ref{Sec:Branching}),   nucleation under a uniform stress in three dimensions (Section \ref{Sec:Uniaxial}), and the thermally driven nucleation of an array of cracks (Section \ref{Sec:Thermal}). Section \ref{Sec:Cost} quantifies the resulting computational saving. We solve each problem with a uniform mesh of size $\hmin$ throughout the domain (labeled \emph{Uniform Refinement}), with adaptive refinement without coarsening (\emph{AMR}), and with adaptive refinement with coarsening (\emph{AMRC}). The three solutions are compared in terms of global responses, crack paths, and crack opening displacements. Global responses in terms of energies and reaction forces are important for understanding whether the elastic energy loss associated with coarsening a phase-field crack into a crack band meaningfully affects the overall solution. Crack-opening displacement results will provide more local insight. All computations reported in this section, except those of Section \ref{Sec:Cost}, were run in parallel on the PACE cluster at the Georgia Institute of Technology. 

Table \ref{tab:params} summarizes the material properties and discretization parameters; unless stated otherwise, $\hmin=\varepsilon/4$, $r=3\varepsilon$, $\theta=0.1$, $v_{\texttt{c}}=0.6$, and $v_a=0.05$. The value of the base mesh size is discussed in Sections \ref{Sec:SENT} and \ref{Sec:Surfing}.

\begin{table}[htbp]
\centering
\caption{Material properties and regularization length used in the benchmark problems. Moduli and strengths are in MPa, $G_c$ in N/mm, and lengths in mm.}
\label{tab:params}
\small
\begin{tabular}{lcccccc}
\toprule
Problem & $E$ & $\nu$ & $G_c$ & $\sts$ & $\scs$ & $\varepsilon$ \\
\midrule
Single-edge notched plate (\ref{Sec:SENT}) & $2.1\times10^{5}$ & 0.3 & 2.7 & 4600 & 23000 & 0.01 \\
Surfing (\ref{Sec:Surfing}) & 9800 & 0.13 & 0.0911 & 27 & 77 & 0.175 \\Echelon cracks (\ref{Sec:Echelon}) & 9800 & 0.13 & 0.0911 & 27 & 77 & 0.3 \\
Uniaxial tension of a rod (\ref{Sec:Uniaxial}) & 70000 & 0.22 & 0.01 & $40\pm5\%$ & 1000 & 0.25 \\
Dynamic branching (\ref{Sec:Branching}) & 72000 & 0.22 & 0.0038 & 6.16 & 18.5 & 0.625 \\
Thermal shock (\ref{Sec:Thermal}) & $3.7\times10^{5}$ & 0.22 & 0.0243 & 240 & 2400 & 0.05 \\
\bottomrule
Problem & $\mu$ & $\Lambda$ & $G_c$ &  $\sts$ & $\shs$  & $\varepsilon$ \\
\midrule
Pure-shear test (\ref{Sec:Slant}) & 0.52 & 85.77 & 0.041 & 0.30 & 1.0  & 0.21 \\
\bottomrule
\end{tabular}
\end{table}

\subsection{Single-edge notched plate under tension}\label{Sec:SENT}

The first problem is the classical single-edge notched plate in tension, shown in Fig.\ \ref{Fig2}(a). It consists of a square plate of side 1 mm with a geometric edge crack of length 0.5 mm at mid-height, subjected to tension under plane strain. The bottom edge is held vertically, and the top edge is displaced vertically by $u_y$ up to $u_0$ = 0.006 mm in 2000 equal increments. Two values are considered for the base mesh size: $h_0=\varepsilon$ and $h_0=2\varepsilon$. The former requires two levels of bisection to reach $\hmin=\varepsilon/4$, while the latter requires three levels.

\begin{figure}[h]
    \centering
    \includegraphics[width=6.5in]{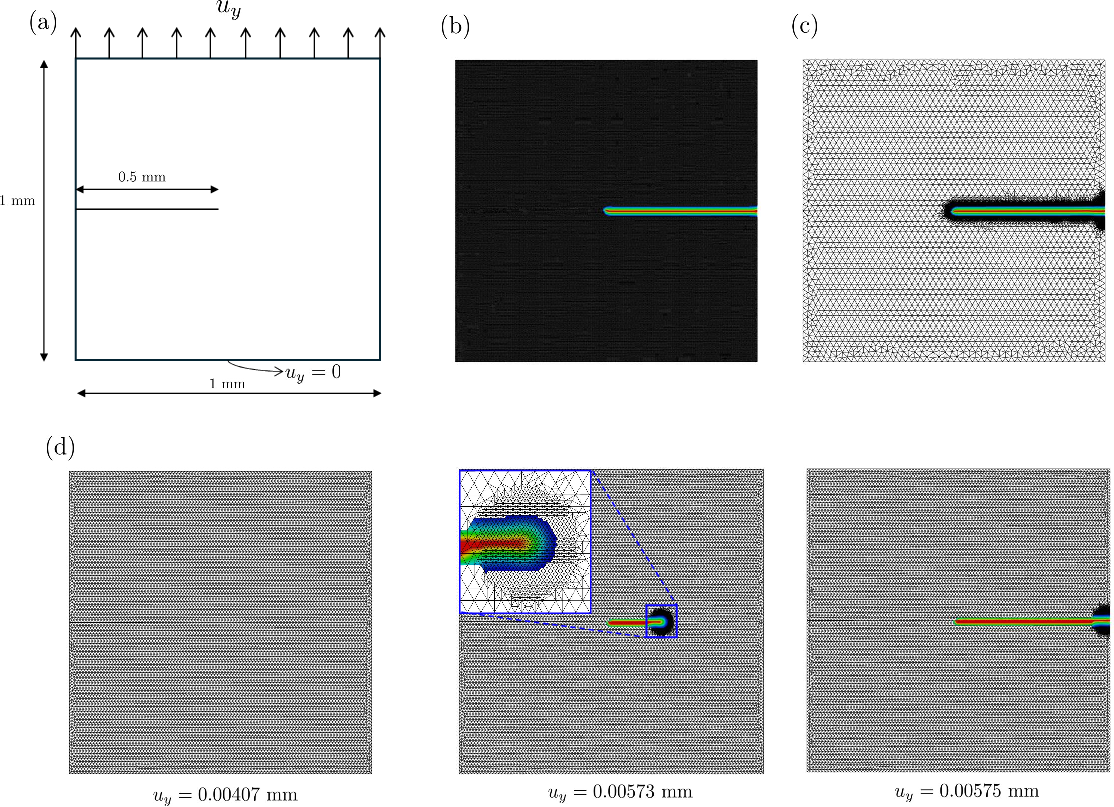}
    \caption{Single-edge notched plate under tension. (a) Geometry and boundary conditions. (b) Phase field at $u_y=0.00575$ mm on the uniform mesh. (c) Phase field and mesh at $u_y=0.00575$ mm of the AMR simulation, in which the mesh remains refined along the entire crack. (d) Snapshots of the AMRC simulation at $u_y=0.00407$, 0.00573, and 0.00575 mm.}\label{Fig2}
\end{figure}

Figure \ref{Fig2}(b)-(c) shows the phase field contours and the meshes at the end of the simulation for the uniform mesh and AMR cases. Fig.~\ref{Fig2}(d) shows the results with AMRC for three different applied displacements. At $u_y=0.00407$ mm, the crack driving force at the notch is still below 90\% of the crack resistance, and the mesh is the base mesh everywhere. At $u_y=0.00573$ mm, the phase-field crack has developed, and the indicator has flagged the process zone at the crack tip. The region surrounding the crack tip is refined, while the region in the wake is coarsened. At $u_y=0.00575$ mm, the crack has run across the plate. In the AMR computation, the mesh remains refined along the whole crack (Fig.\ \ref{Fig2}(c)), while in the AMRC computation, the wake has returned to the base mesh and the only refined region is the small window at the right edge where the last active cells were flagged. The crack in the wake is now represented as a crack band. To understand the effect of the phase-field crack-to-crack-band transition, we next compare the results for the three cases more quantitatively in Fig.~\ref{Fig3}.

The force-displacement curves of Fig.\ \ref{Fig3}(a) obtained with the uniform mesh, with AMR, and with AMRC, for both $h_0=\varepsilon$ and $h_0=2\varepsilon$, are on top: the initial linear response, the peak displacement, and the abrupt drop at fracture coincide. The elastic energy stored in the plate (Fig.\ \ref{Fig3}(b)) coincides as well. Figure \ref{Fig3}(c) provides a more local test for the AMRC formulation. It shows the vertical displacement along the lines $y=\pm2\varepsilon$, that is, along the edges of the regularized crack, relative to the mean displacement $u_0/2$ of the mid-plane, at $u_y=0.00571$ mm, just before the crack starts to advance, and at $u_y=0.00574$ mm, when the tip has reached $x\approx0.9$ mm. The crack advances one time step later with the AMRC formulation; hence, the crack tip is slightly behind the uniform mesh result. 
{Still,} the opening profiles computed across the coarse band of the AMRC mesh are very similar to those of the uniform mesh over the entire length of the crack in magnitude. We will reanalyze the crack opening displacements in the next section where the crack propagation is stable, unlike in this problem.

\begin{figure}[h]
	\centering
	\includegraphics[width=6.5in]{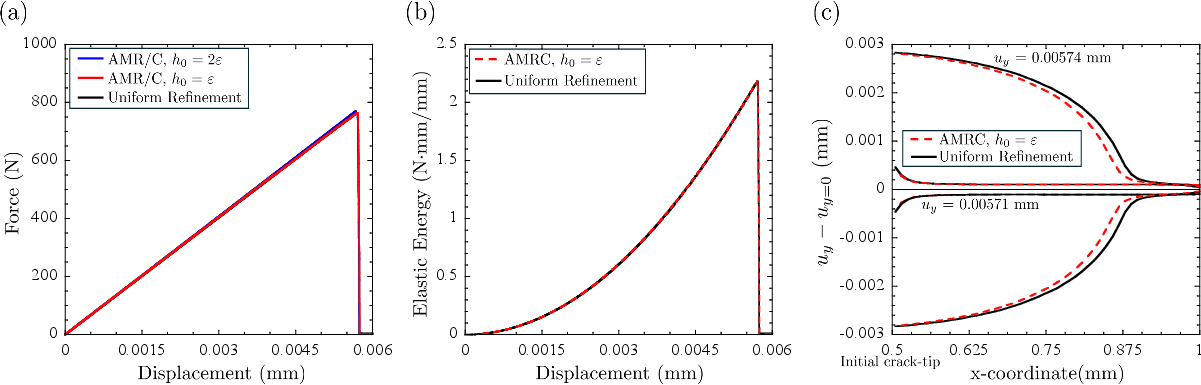}
	\caption{Single-edge notched plate under tension. (a) Force-displacement curves obtained with the uniform mesh and with adaptive refinement with and without coarsening, from base meshes of size $h_0=\varepsilon$ and $h_0=2\varepsilon$. (b) Elastic energy as a function of the applied displacement for the uniform mesh and for AMRC. (c) Vertical displacement along the lines $y=\pm2\varepsilon$ relative to the mid-plane displacement $u_0/2$, at $u_0=0.00571$ mm and $u_0=0.00574$ mm, for the uniform mesh and for AMRC.}\label{Fig3}
\end{figure}

\subsection{The surfing problem}\label{Sec:Surfing}

The surfing problem introduced by Hossain et al.\ \cite{hossain2014} is a standard verification of Griffith-type propagation for phase-field models \cite{Tanne18, KBFLP20, KDK2025Comparison}. A strip of length 30 mm and height 10 mm contains an edge crack of length 5 mm along its mid-plane (Fig.\ \ref{Fig4}(a)). The vertical displacement of the top and bottom edges is prescribed to be that of the mode-I asymptotic solution with stress intensity factor $K_I=\sqrt{EG_c}$, and translated along the direction of the crack at a prescribed velocity of 20 mm/s, over a time $t\in[0,1]$ s discretized in 200 steps. Under this loading, the crack propagates steadily at the same velocity as the loading, and the energy release rate $G$, computed from the $J$-integral along the boundary of the strip, equals $G_c$.

Figure \ref{Fig4}{(a)} shows the AMRC mesh and the phase field contours at $t=0.9$ s. At this stage, the crack has advanced by about 15 mm, and the mesh is the base mesh everywhere except in the small window that travels with the tip. Figure \ref{Fig4}(b) shows the evolution of $G$. For clarity, only the uniform refinement and AMRC cases are shown; the AMR results are on top of the AMRC results. It is observed that for both cases, after the crack initiates from the notch, $G$ settles at $G_c$ for the remainder of the phase-field crack growth. The curves obtained for the two cases are close to each other. This is a global energetic check since if the crack band in the wake does not closely approximate a sharp crack, crack growth will slow down, and this will be reflected as a deviation of $G$ from $G_c$. Figure \ref{Fig4}(c) provides a more local check in terms of the crack opening displacement. It compares the vertical displacement along $y=\pm2\varepsilon$ at $t=0.5$ s and $t=0.9$ s. The opening profiles coincide over the full crack length, most of which lies in the coarsened wake. Therefore, we can conclude that the coarsening has an minimal effect on the solution.  

\begin{figure}[htbp]
	\centering
	\includegraphics[width=6.5in]{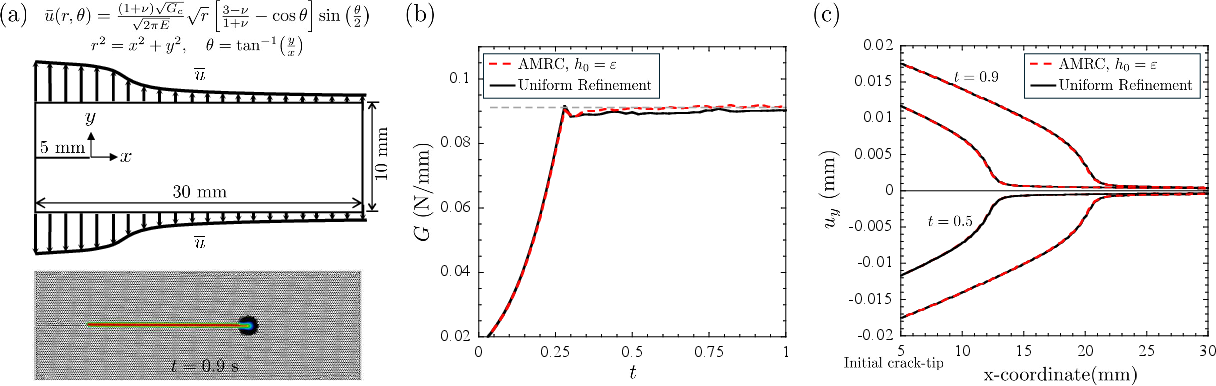}
	\caption{The surfing problem. (a) Geometry and boundary conditions, and the AMRC mesh and phase field at $t=0.9$ s. (b) Energy release rate computed from the $J$-integral as a function of time for the uniform mesh and for AMRC; the dashed line is $G_c$. (c) Vertical displacement along the lines $y=\pm2\varepsilon$ at $t=0.5$ s and $t=0.9$ s for the uniform mesh and for AMRC.}\label{Fig4}
\end{figure}

{Next we investigate the effect of mesh size in the coarsened crack region.
We repeat the AMRC computation with base meshes of size $h_0=2\varepsilon$, $4\varepsilon$, and $8\varepsilon$, in every case refined to the same target size $\hmin=\varepsilon/4$ around the tip. Note that the phase field crack width is $4\varepsilon$ in the fine region with AT$_1$ regularization. Results are shown in Fig.\ \ref{Fig4_1}.  For $h_0=\varepsilon$ and $2\varepsilon$, the coarse crack band remains a straight band (Fig.\ \ref{Fig4_1}(a)) and the energy release rate stays approximately at $G_c$ (Fig.\ \ref{Fig4_1}(b)). For $h_0=4\varepsilon$ and $h_0=8\varepsilon$, $G$ significantly overshoots $G_c$ and oscillates, and the crack band becomes jagged. The energy release rate shows increasing error as the band size is widened and the crack path becomes jagged. Therefore, we conclude that for the current coarsening algorithm to produce reasonable results, the base mesh must satisfy $h_0\le2\varepsilon$. Using remeshing to align the jagged crack band correctly will likely improve the results for higher values of $h_0$.
Note that $h_0=2\varepsilon$ with target mesh size of $\varepsilon/4$ represents a reduction in mesh size in the crack wake by a factor of 64 for 2D problems and 512 for 3D problems.
}

\begin{figure}[h]
	\centering
	\includegraphics[width=6in]{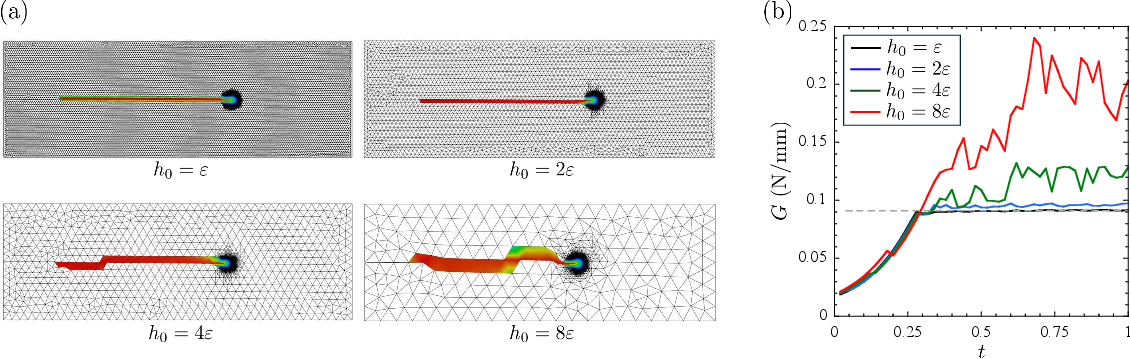}
	\caption{{The surfing problem with coarser base meshes. (a) Phase field and mesh at $t=0.9$ s for base meshes of size $h_0=\varepsilon$, $2\varepsilon$, $4\varepsilon$, and $8\varepsilon$, all refined to $\hmin=\varepsilon/4$ around the crack tip. (b) Energy release rate computed from the $J$-integral as a function of time for the four base meshes; the dashed line is $G_c$.}}\label{Fig4_1}
\end{figure}

\subsection{Echelon cracks under anti-plane shear}\label{Sec:Echelon}
We next evaluate our adaptive formulation for a complex three-dimensional crack growth problem where the crack front breaks up into an array of inclined cracks, known as echelon cracks, under dominant mode III loading. This phenomenon is observed in experiments on a variety of brittle materials \cite{lazarus2008comparison, pham2017echelon, chen2015crack3d}. Ward and Kumar \cite{WK2025} recently showed that the phase-field model of Section \ref{Sec:Theory} predicts the fragmentation as a nucleation event governed by the strength surface and the fracture length scale, without input of geometric or material disorder.
We revisit their problem and consider a cube of side 25 mm made of graphite, containing a planar crack of length 12.5 mm (Fig.\ \ref{Fig6}(a)). The bottom face is fixed, and the top face is displaced along anti-plane direction, $u_z=\bar{u}_z$, so that the front is loaded in mode III.

Figure \ref{Fig6}(b) shows the daughter cracks predicted by the AMRC simulation in comparison to the results from the uniform refinement case as seen on the $y$-$z$ and $x$-$z$ planes. Both cases show an array of inclined segments nucleated along the front that closely resemble each other in terms of the number of segments, their positions along the front, and their sizes. The force-displacement response (Fig.\ \ref{Fig6}(c)) coincides as well. In the AMRC simulation, each daughter crack nucleates within its own refinement window that is coarsened as the crack evolves, so this comparison provides a good indication of the performance of the AMRC formulation when multiple 3D cracks are involved.

\begin{figure}[htbp]
	\centering
	\includegraphics[width=6in]{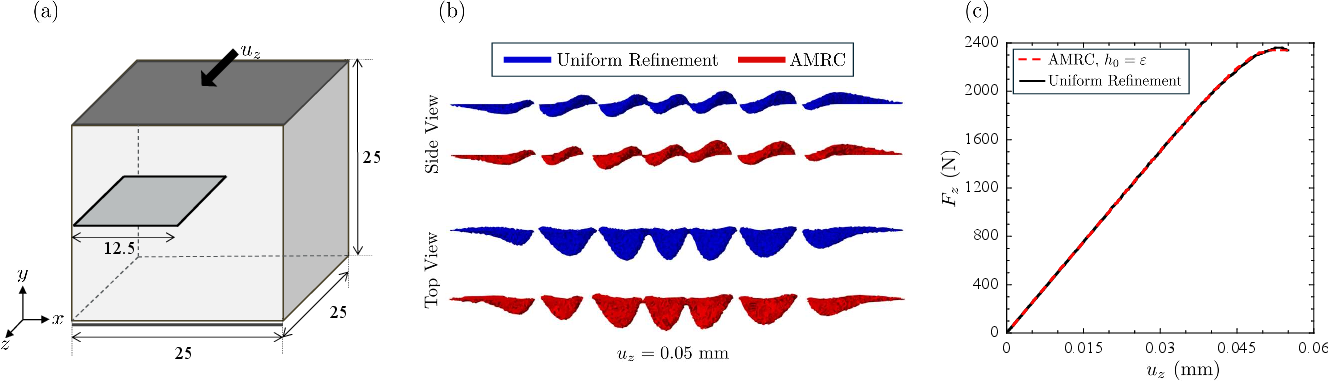}
	\caption{Echelon cracks under anti-plane shear. (a) Geometry and boundary conditions. (b) Daughter cracks {as seen on the $y$-$z$ plane (side view) and the $x$-$z$ plane (top view)} at $u_z=0.05$ mm, for the reference mesh (blue) and for AMRC (red). (c) Force-displacement curves of the top face for the reference mesh and for AMRC.}\label{Fig6}
\end{figure}

\begin{figure}[h]
	\centering
	\includegraphics[width=5.in]{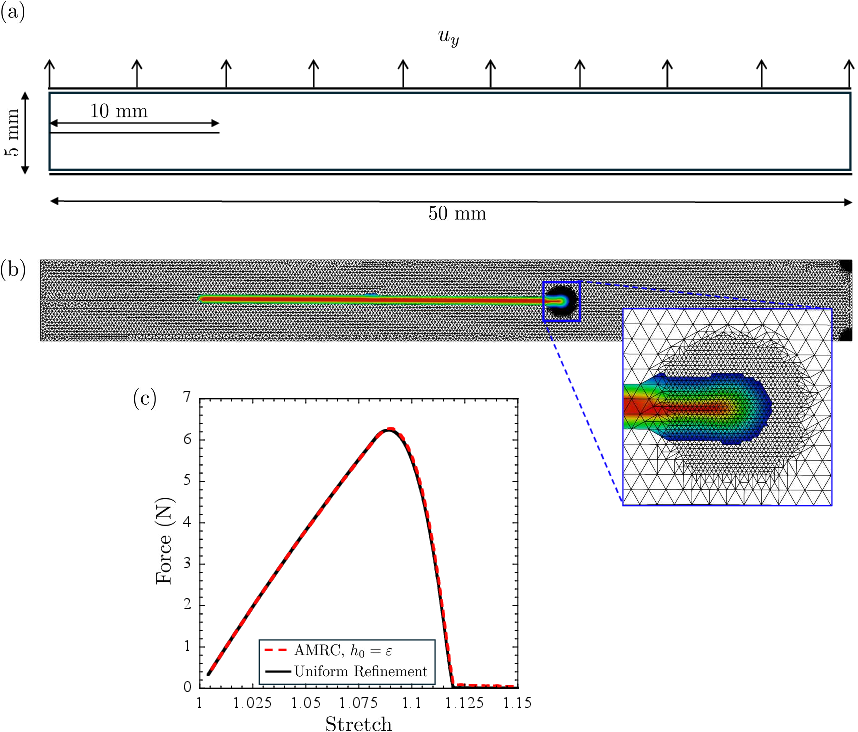}
	\caption{{Pure-shear fracture test of a polyurethane elastomer. (a) Geometry and boundary conditions. (b) Phase field at a stretch $\lambda=1.1$ on the AMRC mesh; the inset shows the refined window at the crack tip. (c) Force on the top grip as a function of the stretch for the uniform mesh and for AMRC.}}\label{Fig7}
\end{figure}

\subsection{{Pure-shear fracture test of an elastomer}}\label{Sec:Slant}
{
The next problem combines crack propagation with finite deformations and near incompressibility. It is the pure-shear fracture test, the fourth of the challenge problems of Kamarei et al.\ \cite{kamarei2026nine}, and a standard experimental and numerical problem to study crack growth in soft materials. A thin strip of a polyurethane elastomer is bonded along its top and bottom edges to rigid grips and contains an edge crack along its mid-plane (Fig.\ \ref{Fig7}(a)). The bottom grip is fixed, and the top grip is displaced vertically with its horizontal displacement suppressed, so that the separation $h$ of the grips prescribes the stretch $\lambda=h/H$ on the strip of height $H$. 
Neo-Hookean material response is assumed, and the material constants are listed in Table \ref{tab:params}. The computation is carried out under plane stress. The displacement field is discretized with the stabilized linear Crouzeix-Raviart element \cite{KFLP18}, which has about three times as many degrees of freedom as the linear Lagrange element on the same mesh, so that the savings brought by AMRC are correspondingly larger.

Figure \ref{Fig7}(b) shows the phase field at $\lambda=1.1$, when the crack has run across more than half of the strip on the AMRC mesh. As in the previous examples, the AMRC mesh is the base mesh everywhere except in the window that travels with the tip (inset), and the wake is carried by the coarse band. The force-stretch curves obtained with the uniform mesh and with AMRC coincide, both in the critical stretch and in the rate at which the force drops.  The critical stretch $\lambda_c \approx1.09$ is also in agreement with the value $\lambda_c=1.093$ obtained from the analytical formula $H\,W(\lambda_c)=G_c$.
}

\subsection{Dynamic crack branching}\label{Sec:Branching}

The next problem tests the framework under dynamic loading and where a crack branches, so that the active set and the refined window split. We consider the crack-branching benchmark of Fig.\ \ref{Fig5}(a) \cite{Song2008_ComparativeFE, borden2012phase, Bleyer2017_DynamicCrackVariational, dahal2026dynamic}: a thin plate of 100 mm by 40 mm with an edge crack of length 50 mm along its mid-plane, loaded by a tensile traction of 2 MPa applied on the top and bottom edges within a single time step of 25 ns and then held constant. The material is glass with properties listed in Table \ref{tab:params}.

The contour plots of the phase field are shown in Fig.\ \ref{Fig5}(b)-(e) for uniform mesh, AMR mesh, and AMRC mesh.
In all cases, the crack starts to propagate shortly after the stress waves reach the tip, accelerates along the mid-plane, and branches at roughly two-thirds of the plate length into two branches that reach the right edge of the plate. The final crack pattern on the uniform mesh (Fig.\ \ref{Fig5}(b)), the AMR mesh (Fig.\ \ref{Fig5}(c)) and  the AMRC mesh (Fig.\ \ref{Fig5}(e)) are the same. The AMRC snapshots at $t=23$ $\mu$s and $t=37$ $\mu$s (Fig.\ \ref{Fig5}(d,e)) show a single window traveling with the tip before branching and two windows following the two branches afterward, with the entire wake, including the branching point, represented by the base mesh. Branching required no additional intervention in the numerical implementation. Figure \ref{Fig5}(f) compares the elastic and kinetic energies of the plate as functions of time for the uniform mesh and for AMRC. Both energies match throughout the simulation, and they also match with the AMR results (not included). 

\begin{figure}[h]
	\centering
	\includegraphics[width=6.5in]{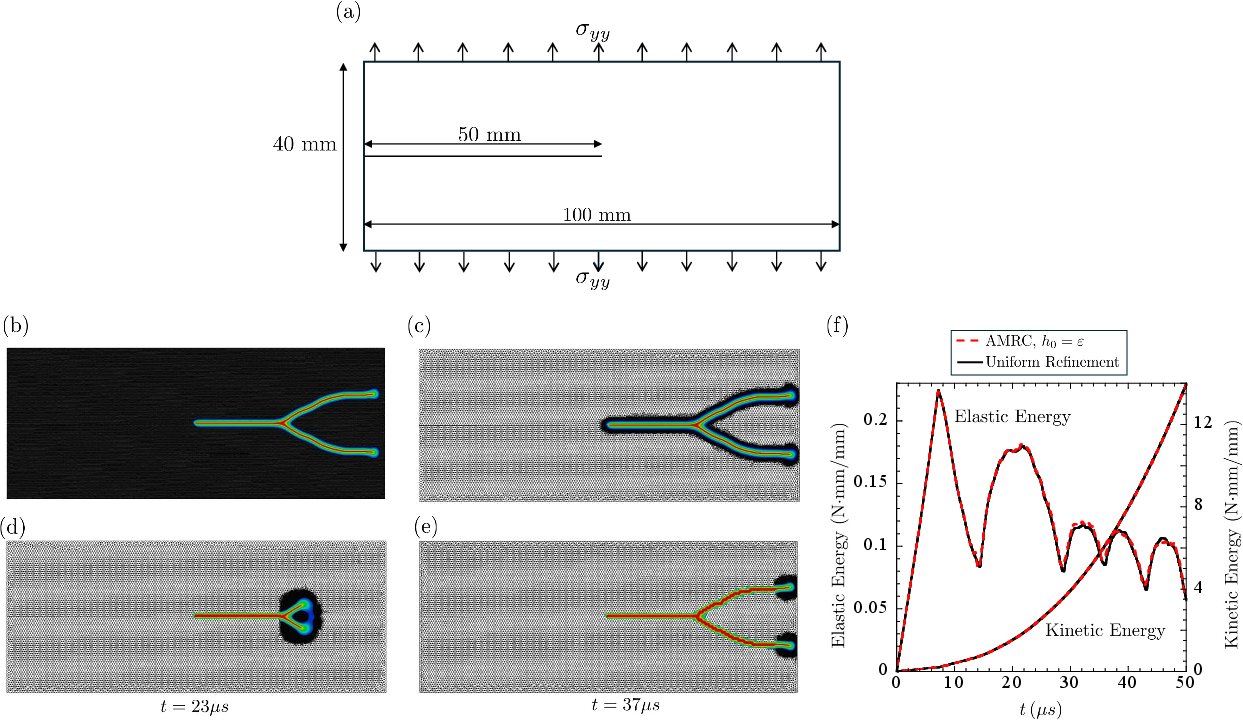}
	\caption{Dynamic crack branching. (a) Geometry and loading. (b) Final crack pattern on the uniform mesh. (c) Final crack pattern and mesh of the AMR simulation. (d, e) Snapshots of the AMRC simulation at $t=23$ $\mu$s, during branching, and $t=37$ $\mu$s, after branching, showing the refined windows traveling with the crack tips. (f) Elastic and kinetic energies as functions of time for the uniform mesh and for AMRC.}\label{Fig5}
\end{figure}

\begin{figure}[h]
	\centering
	\includegraphics[width=5.in]{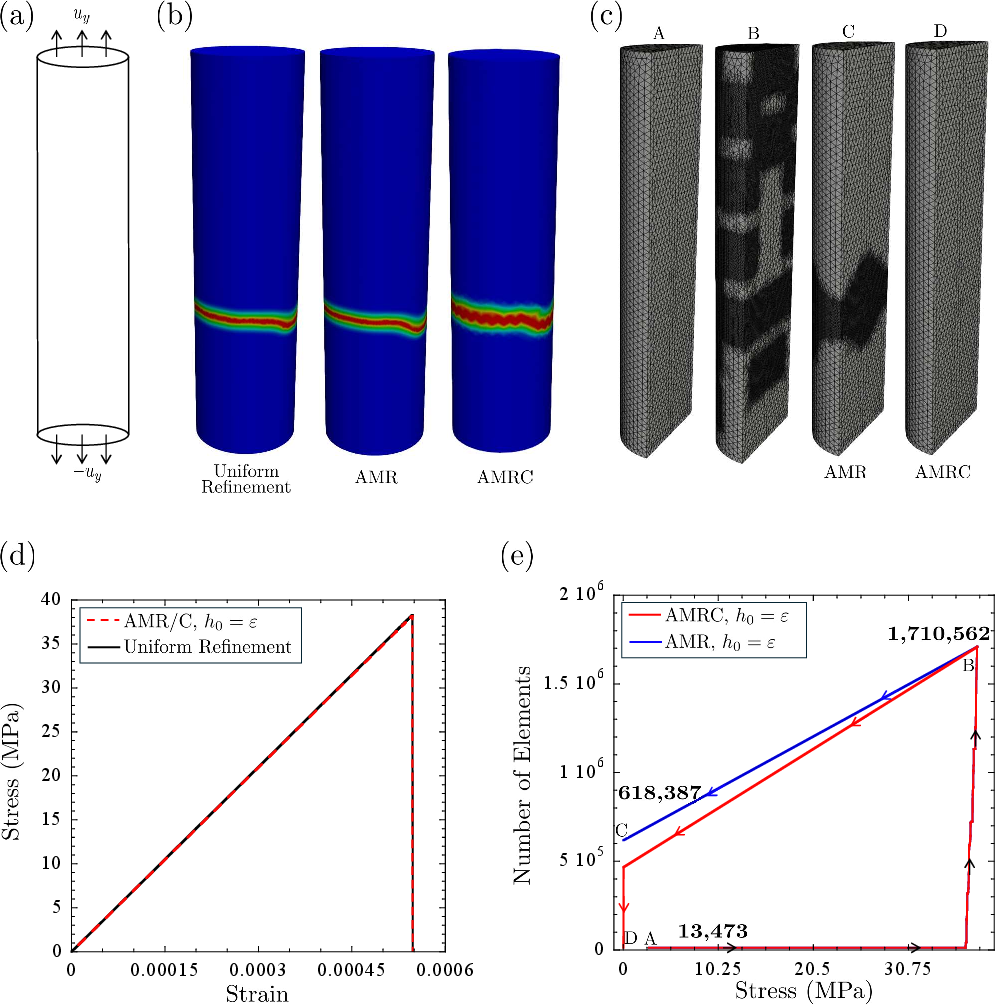}
	\caption{Uniaxial tension of a glass rod with a randomly perturbed tensile strength. (a) Geometry and loading. (b) Crack predicted with the uniform mesh (left), with AMR (middle), and with AMRC (right). (c) AMR/C mesh at the start of the simulation, immediately before nucleation, after fracture for AMR and after fracture for AMRC. (d) Stress-strain curves for the uniform mesh and for AMR/C. (e) Number of elements in the AMR and AMRC mesh as a function of the stress; the arrows indicate the direction of time.}\label{Fig8}
\end{figure} 

\subsection{Nucleation under uniform stress: uniaxial tension of a glass rod}\label{Sec:Uniaxial}

Nucleation under spatially uniform stress is a demanding case for any adaptive strategy, because the fields cannot distinguish the exact location of the future crack until it localizes, even with the use of stochasticity. We consider the uniaxial tension test of a soda-lime glass rod from the first challenge problem of Kamarei et al.\ \cite{kamarei2026nine}. A rod of length 15 mm and radius 2 mm is stretched by axial displacements $\pm u_y$ applied at its ends (Fig.\ \ref{Fig8}(a)). Only one quarter of the rod is modeled using symmetry. Following \cite{kamarei2026nine}, the tensile strength is perturbed by $\pm5\%$ on random cubic subdomains of size $5\varepsilon$ to break the uniformity of the fields. The perturbation is defined as a deterministic function of position so that it is identical on every mesh.

Results are shown in Fig.\ \ref{Fig8}(b)-(e) for the uniform mesh, AMR mesh, and AMRC mesh.
The stress-strain response (Fig.\ \ref{Fig8}(d)) is linear up to 38 MPa, the strength of the weakest subdomains, and then drops abruptly as a severing crack nucleates perpendicular to the axis (Fig.\ \ref{Fig8}(b)). The AMRC and uniform-mesh responses coincide, as do the crack locations and shapes. Figure \ref{Fig8}(c) shows the evolution of the AMR and the AMRC mesh. During the elastic loading, up to about 95\% of the strength, the mesh is the base mesh. As the stress approaches the strength, the indicator flags the subdomains in which the strength is lowest, which appear as patches spread over the entire domain in the second panel of Fig.\ \ref{Fig8}(c). The number of elements (Fig.\ \ref{Fig8}(e)) rises from that of the base mesh to about $1.7\times10^{6}$ within a few steps, which is comparable to the uniform mesh. Once the crack forms and the stress drops, the active set shrinks with the stress. Each remeshing refines a smaller region, and the mesh coarsens progressively until only the surroundings of the crack remain refined (third panel of Fig.\ \ref{Fig8}(c)) for AMR, whereas the mesh returns to the base mesh in AMRC (last panel of Fig.\ \ref{Fig8}(c)). The fine mesh is thus needed only during the few steps in which nucleation takes place. Unlike criteria based on the phase field or strain energy, the strength-surface indicator produces the fine mesh before the phase field starts to evolve. Nucleation is therefore correctly resolved at $\hmin$, and the predicted strength is that of the uniform mesh.

\subsection{Thermal shock of a ceramic slab}\label{Sec:Thermal}

The last problem is the quenching of a hot ceramic slab, in which a periodic array of cracks nucleates at the quenched surface. It has been studied experimentally by Shao et al.\ \cite{shao2011quenching} and Jiang et al.\ \cite{jiang2012thermal}, analyzed with gradient damage models by Sicsic et al.\ \cite{sicsic2014thermal} and Bourdin et al.\ \cite{bourdin2014morphogenesis}, and recently revisited with the complete phase-field model by Zeng and Dolbow \cite{zengdolbow2026thermal}. An alumina slab of 50 mm by 10 mm, initially at $T_0=300^\circ$C, is immersed in a water bath at $20^\circ$C. Using symmetry, one quarter of the slab, 25 mm by 5 mm, is modeled (Fig.\ \ref{Fig9}(a)), with convective heat exchange with the bath on the two free edges characterized by a heat transfer coefficient of $54.5$ kW/(m$^2$K). The temperature-dependent conductivity, specific heat, and thermal expansion coefficient are taken from \cite{jiang2012thermal}, under plane stress. As in \cite{dolbow2025uniform}, the tensile strength is a random field.

As the surface layer cools and contracts, tensile stress builds up along the quenched edges. The indicator first flags the entire boundary layer, then refines it, and an array of cracks nucleates in it with spacing controlled by thermal stresses and fracture properties. Fig.\ \ref{Fig9}(b)-(d) show that the three discretizations predict the same pattern, including the same number of cracks, the same spacing, the same arrest lengths, and the same deflections of the growing cracks. The AMR mesh (Fig.\ \ref{Fig9}(c)) remains refined along the entire length of the cracks. In the AMRC mesh (Fig.\ \ref{Fig9}(d)), only the tips of the cracks  are surrounded by a refined window.
The wakes of all the cracks are carried by the base mesh. Together with the results of the last problem, these results show that the AMRC formulation seamlessly handles complex nucleation problems.

\begin{figure}[h]
	\centering
	\includegraphics[width=6.5in]{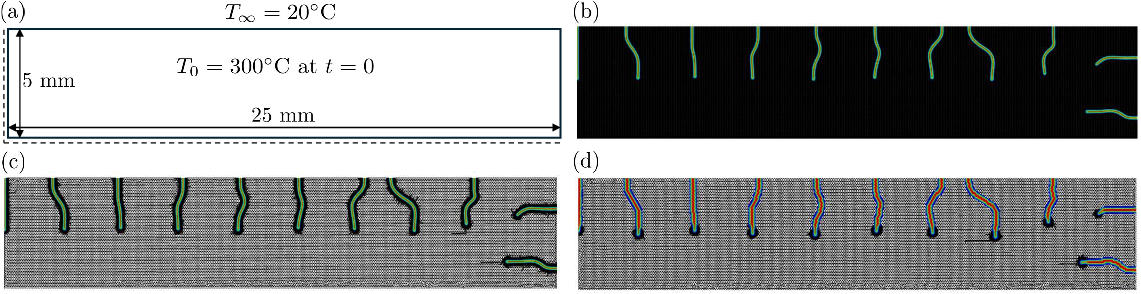}
	\caption{Thermal shock of a ceramic slab. (a) Geometry and thermal boundary conditions of the quarter model; the dashed lines are the symmetry planes. Final crack patterns predicted with (b) the uniform mesh, (c) AMR, and (d) AMRC.}\label{Fig9}
\end{figure}

\subsection{Problem size and computational cost}\label{Sec:Cost}
{
The benefit of coarsening the crack wake is best seen in the evolution of the number of elements during a simulation. Figure \ref{Fig10}(a) shows this evolution for the surfing problem of Section \ref{Sec:Surfing}. The uniform mesh has 364,265 elements. The AMR mesh starts with the 24,594 elements of the base mesh and, once the crack starts to propagate at $t\approx0.27$ s, grows in steps corresponding to each remeshing event. As the refined region lengthens with the crack, the number of elements reaches 43,079 at $t=1$ s and would keep growing in proportion to the crack length. The AMRC mesh, by contrast, stays within a few percent of the base-mesh count throughout. The refined window simply travels with the tip, and each remeshing event removes as many elements from the wake as it adds ahead of the front. 
The same behavior is observed in all the propagation problems shown above.

The wall times of the three strategies were measured on a single core of a desktop workstation with an Intel Core i9-13900K processor and 32 GB of memory, and are shown in Fig.\ \ref{Fig10}(b). The staggered iterations are left uncapped, and every step converged to the prescribed tolerance \texttt{tol} = $1e^{-7}$. The uniform-mesh simulation took 24,256 s, the AMR simulation 1,872 s, and the AMRC simulation 1,233 s, that is, 13 and 20 times faster than the uniform mesh, respectively. 
The additional savings of AMRC with respect to AMR grow with the amount of crack surface created during the simulation. For cracks that slowly grow to a long area or advance over many loading cycles, the problem size of AMR keeps growing while that of AMRC remains fixed. Note that in our simulations, we have tied the crack band mesh size to the base mesh size for simplicity of comparison. Uncoupling the two would bring more savings since generally the mesh size far away from the cracks need not be  $\mathcal{O}(\varepsilon)$.

}

\begin{figure}[h]
	\centering
	\includegraphics[width=5.5in]{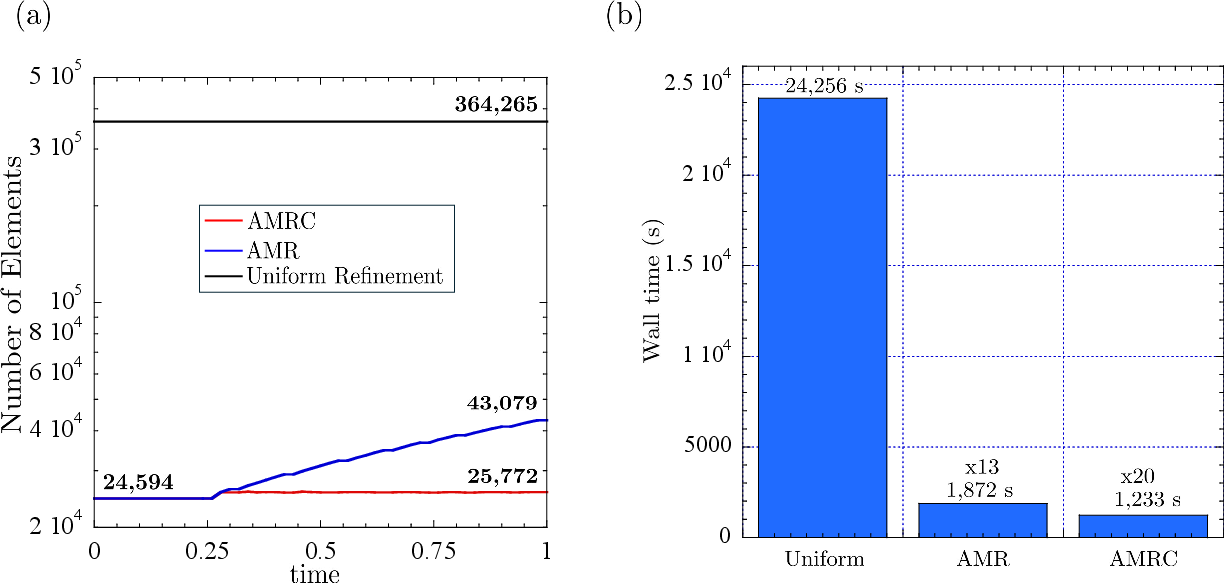}
	\caption{Computational cost for the surfing problem. (a) Number of elements as a function of time for the uniform mesh, for adaptive refinement without coarsening, and for adaptive refinement with coarsening. (b) Wall time of the three simulations.}\label{Fig10}
\end{figure}

\section{Final comments}\label{Sec:Conclusions}

The adaptive mesh refinement and coarsening framework for phase-field fracture presented in this work is based on two principles. First, the framework should be based on a physical indicator for crack evolution so it can be used universally for any material, loading, and phase-field formulation. Second, the implementation should be simple so that it can be easily used with different finite element libraries and software. We achieved the former by basing the indicator on material strength, since violation of the strength surface is a necessary condition for the phase field to evolve. We demonstrated its application to seven benchmark problems, spanning quasi-static and dynamic propagation, crack branching, three-dimensional crack front segmentation, finite deformations in a nearly incompressible material, nucleation under uniform stress, and thermal shock. We showed that the refinement and coarsening algorithm performs robustly throughout. We achieved the latter by devising a new coarsening strategy based on refinement from a fixed base mesh. What's more, we demonstrated that replacing the crack wake with a coarse crack band has minimal effect on the global and local response and can be used as an effective strategy to keep the fine mesh confined to a small region near the tip. 

With the objective of making the formulation as accessible as possible, we have made available on GitHub the implementation in FEniCSx for all the problems studied in this work. The implementation runs in parallel and can be used with any phase field method with minimal changes. The algorithm is also relatively straightforward to implement in other finite element libraries such as MFEM and MOOSE.

{
The next major step in making phase field simulations competitive for large structural simulations, while providing more accurate physical information than competing methods, is to reduce the number of staggered iterations required for convergence at each step. Typically, the staggered fixed-point iteration converges very slowly and can take hundreds, if not thousands, of iterations to converge to a low tolerance. A very low tolerance is not essential for most problems with stable crack growth---a good rule of thumb is to cap iterations at 200. Even then, the cost of each step is staggering. Therefore, we need to design tailored acceleration methods for fixed-point iteration. We show an example below of how acceleration methods can reduce the cost.

We adopt Anderson acceleration (AA) \cite{anderson1965iterative, walker2011anderson} of the staggered scheme---used for phase-field fracture by Storvik et al.\ \cite{storvik2021accelerated}---and re-solve the surfing problem. Crack growth is not abrupt in the surfing problem, so applying AA should be stable. The staggered iteration of Algorithm \ref{alg:step} is a fixed-point iteration $x^{(i)}=S(x^{(i-1)})$ for the pair $x=(\bfu,v)$. AA replaces each new iterate with the linear combination of the last $m+1$ iterates that minimizes the norm of the fixed-point residual $S(x)-x$ in the least-squares sense. We use a depth $m=3$. Following Storvik et al., the acceleration is applied only while the displacement residual decreases from one iteration to the next. Otherwise, we use plain staggered iterations until five consecutive residuals decrease, after which we restart the acceleration. The accelerated phase field is projected onto the bounds $[v_{\rm lb},v_{\rm ub}]$, the state accepted at the end of a step is always the output of the last solve of the phase-field problem, and the acceleration is restarted after each remeshing, since the stored iterates live on the old mesh.

Figure \ref{Fig11}(a) shows that the energy release rates obtained with Anderson acceleration applied on uniform mesh and on AMRC mesh and shows that it does not affect the accuracy of the solution. Figure \ref{Fig11}(b) compares the wall times. Anderson acceleration alone, on the uniform mesh, reduces the wall time of 24,256 s by a factor of 6, which reflects the reduction in the number of staggered iterations to around 50. Combined with AMRC, whose own factor is 20, it yields savings by a factor of 119, reducing the simulation time to less than 4 minutes. We believe that even greater savings can be achieved by developing more tailored methods rather than using off-the-shelf methods. An ultimate goal should be to reduce the number of staggered iterations necessary to converge to a satisfactorily low tolerance by another order of magnitude.

Less prominently, we can gain additional savings by adopting adaptive temporal refinement and coarsening (ATRC), especially for smaller problems. We show here the impact on the surfing problem of using the ATRC method of Rohracker et al.\ \cite{rohracker2026efficient}, which adapts the load step to the rate of crack growth. The increment of the regularized fracture energy $\mathcal{G}=\int_\Omega\frac{3G_c}{8}\left(\frac{1-v}{\varepsilon}+\varepsilon|\nabla v|^2\right)\,{\rm d}\bfX$ is monitored during the staggered iterations of every step. When it exceeds the increment of the previous step by more than a tolerance, set to half the increment produced by steady propagation during one fixed step of Section \ref{Sec:Surfing}, the step is rejected and repeated with a four times smaller increment, down to two levels below the initial one; the step size is doubled again once the interval covered by the last coarse step has been completed without triggering the criterion.

Figure \ref{Fig11}(a) also shows that the energy release rate obtained with ATRC combined with AMRC and AA is accurate.  Figure \ref{Fig11}(b) shows that the wall time further drops to just 74 s. We will explore and incorporate these and other methods, such as the recent acceleration scheme proposed by Shala and Waisman \cite{waisman2026acceleration}, in future work.

}

\begin{figure}[h]
	\centering
	\includegraphics[width=5.5in]{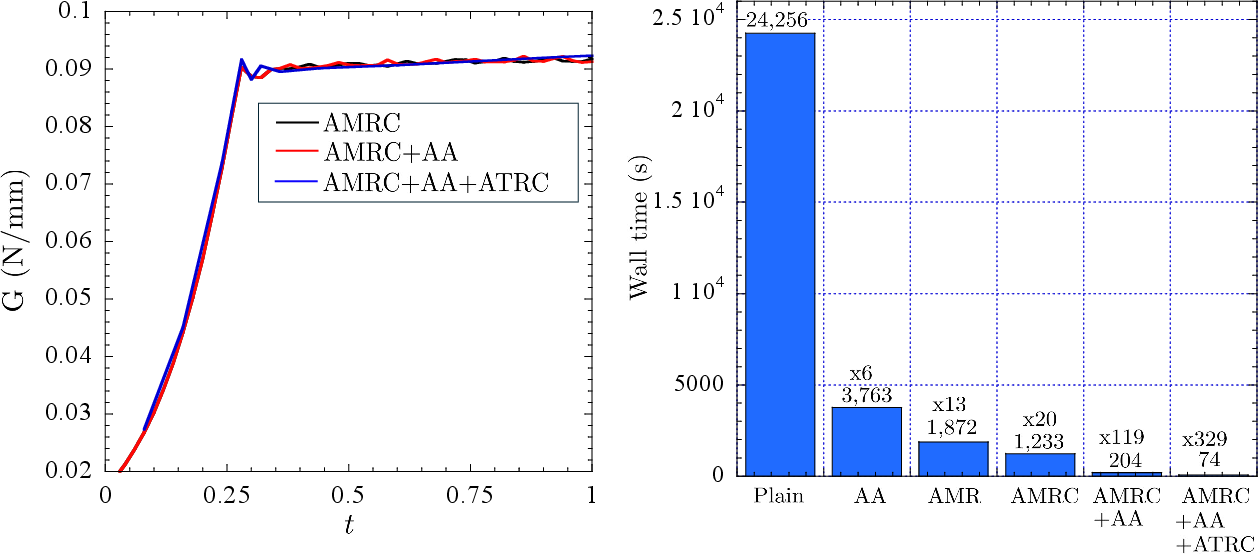}
	\caption{{Acceleration of the surfing problem. (a) Energy release rate as a function of time for AMRC, for AMRC with Anderson acceleration (AA), and for AMRC with Anderson acceleration and adaptive temporal refinement and coarsening (ATRC). (b) Wall time of the uniform-mesh simulation (Plain), of Anderson acceleration on the uniform mesh (AA), of AMR, of AMRC, and of the combinations of AMRC with AA and with AA and ATRC; the factors are the speed-ups relative to the uniform mesh.}}\label{Fig11}
\end{figure}

\section*{Acknowledgements}

\noindent AK and AD would like to acknowledge the financial support from the National Science Foundation, United States, through the grants CMMI-2404808 and 2541441.  

\section*{Data Availability}
The entire source code is available at \url{https://github.com/Aditya-Kumar-Lab-GT/AMRC}.

\bibliographystyle{unsrt}
\bibliography{main}

\begin{thebibliography}{10}

\bibitem{ErdoganSih1963}
Fazil Erdogan and GC~Sih.
\newblock On the crack extension in plates under plane loading and transverse shear.
\newblock {\em Journal of basic engineering}, 85(4):519--525, 1963.

\bibitem{sih1974strain}
George~C Sih.
\newblock Strain-energy-density factor applied to mixed mode crack problems.
\newblock {\em International Journal of fracture}, 10(3):305--321, 1974.

\bibitem{goldstein1974brittle}
Robert~V Gol'dstein and Rafael~L Salganik.
\newblock Brittle fracture of solids with arbitrary cracks.
\newblock {\em International Journal of Fracture}, 10(4):507--523, 1974.

\bibitem{nuismer1975energyrelease}
RJ~Nuismer.
\newblock An energy release rate criterion for mixed mode fracture.
\newblock {\em International Journal of Fracture}, 11(2):245--250, 1975.

\bibitem{wu1978fracture}
C-H Wu.
\newblock Fracture under combined loads by maximum-energy-release-rate criterion.
\newblock {\em Journal of Applied Mechanics}, 45(3):553, 1978.

\bibitem{Francfort98}
Gilles~A Francfort and J-J Marigo.
\newblock Revisiting brittle fracture as an energy minimization problem.
\newblock {\em Journal of the Mechanics and Physics of Solids}, 46(8):1319--1342, 1998.

\bibitem{Bourdin00}
Blaise Bourdin, Gilles~A Francfort, and Jean-Jacques Marigo.
\newblock Numerical experiments in revisited brittle fracture.
\newblock {\em Journal of the Mechanics and Physics of Solids}, 48(4):797--826, 2000.

\bibitem{WK2025}
Olivia Ward and Aditya Kumar.
\newblock Why planar cracks fragment into echelon cracks.
\newblock {\em Extreme Mechanics Letters}, 87:102504, 2026.

\bibitem{borden2012phase}
Michael~J. Borden, Clemens~V. Verhoosel, Michael~A. Scott, Thomas~J.R. Hughes, and Chad~M. Landis.
\newblock A phase-field description of dynamic brittle fracture.
\newblock {\em Computer Methods in Applied Mechanics and Engineering}, 217--220:77--95, 2012.
\newblock Available at: \url{www.elsevier.com/locate/cma}.

\bibitem{borden2016phase}
Michael~J Borden, Thomas~JR Hughes, Chad~M Landis, Amin Anvari, and Isaac~J Lee.
\newblock A phase-field formulation for fracture in ductile materials: Finite deformation balance law derivation, plastic degradation, and stress triaxiality effects.
\newblock {\em Computer Methods in Applied Mechanics and Engineering}, 312:130--166, 2016.

\bibitem{bourdin2014morphogenesis}
Blaise Bourdin, Jean-Jacques Marigo, Corrado Maurini, and Paul Sicsic.
\newblock Morphogenesis and propagation of complex cracks induced by thermal shocks.
\newblock {\em Physical Review Letters}, 112(1):014301, 2014.

\bibitem{Tanne18}
Erwan Tann{\'e}, Tianyi Li, Blaise Bourdin, Jean-Jacques Marigo, and Corrado Maurini.
\newblock Crack nucleation in variational phase-field models of brittle fracture.
\newblock {\em Journal of the Mechanics and Physics of Solids}, 110:80--99, 2018.

\bibitem{duarte2020validation}
Faisal~M Mukhtar, Phillipe~D Alves, and C~Armando Duarte.
\newblock Validation of a 3-d adaptive stable generalized/extended finite element method for mixed-mode brittle fracture propagation.
\newblock {\em International Journal of Fracture}, 225(2):129--152, 2020.

\bibitem{KFLP18}
Aditya Kumar, Gilles~A Francfort, and Oscar Lopez-Pamies.
\newblock Fracture and healing of elastomers: A phase-transition theory and numerical implementation.
\newblock {\em Journal of the Mechanics and Physics of Solids}, 112:523--551, 2018.

\bibitem{talamini2021attaining}
Brandon Talamini, Michael~R Tupek, Andrew~J Stershic, Tianchen Hu, James~W Foulk~III, Jakob~T Ostien, and John~E Dolbow.
\newblock Attaining regularization length insensitivity in phase-field models of ductile failure.
\newblock {\em Computer Methods in Applied Mechanics and Engineering}, 384:113936, 2021.

\bibitem{pham2011gradient}
Kim Pham, Hanen Amor, Jean-Jacques Marigo, and Corrado Maurini.
\newblock Gradient damage models and their use to approximate brittle fracture.
\newblock {\em International Journal of Damage Mechanics}, 20(4):618--652, 2011.

\bibitem{KLP20}
Aditya Kumar and Oscar Lopez-Pamies.
\newblock The phase-field approach to self-healable fracture of elastomers: A model accounting for fracture nucleation at large, with application to a class of conspicuous experiments.
\newblock {\em Theoretical and Applied Fracture Mechanics}, 107:102550, 2020.

\bibitem{KRLP22}
Aditya Kumar, K~Ravi-Chandar, and Oscar Lopez-Pamies.
\newblock The revisited phase-field approach to brittle fracture: application to indentation and notch problems.
\newblock {\em International Journal of Fracture}, 237(1-2):83--100, 2022.

\bibitem{KBFLP20}
Aditya Kumar, Blaise Bourdin, Gilles~A Francfort, and Oscar Lopez-Pamies.
\newblock Revisiting nucleation in the phase-field approach to brittle fracture.
\newblock {\em Journal of the Mechanics and Physics of Solids}, 142:104027, 2020.

\bibitem{Wu18}
Jian-Ying Wu and Vinh~Phu Nguyen.
\newblock A length scale insensitive phase-field damage model for brittle fracture.
\newblock {\em Journal of the Mechanics and Physics of Solids}, 119:20--42, 2018.

\bibitem{heister2015prisms}
Timo Heister, Mary~F Wheeler, and Thomas Wick.
\newblock A primal-dual active set method and predictor-corrector mesh adaptivity for computing fracture propagation using a phase-field approach.
\newblock {\em Computer Methods in Applied Mechanics and Engineering}, 290:466--495, 2015.

\bibitem{Gupta2022_AdaptiveMeshRefinement}
Abhinav Gupta, U.~Meenu Krishnan, Tushar~Kanti Mandal, Rajib Chowdhury, and Vinh~Phu Nguyen.
\newblock An adaptive mesh refinement algorithm for phase-field fracture models: Application to brittle, cohesive, and dynamic fracture.
\newblock {\em Computer Methods in Applied Mechanics and Engineering}, 399:115347, 2022.

\bibitem{hirshikesh2021adaptive}
H~Hirshikesh, ALN Pramod, Haim Waisman, and S~Natarajan.
\newblock Adaptive phase field method using novel physics based refinement criteria.
\newblock {\em Computer Methods in Applied Mechanics and Engineering}, 383:113874, 2021.

\bibitem{wick2016goal}
Thomas Wick.
\newblock Goal functional evaluations for phase-field fracture using {PU}-based {DWR} mesh adaptivity.
\newblock {\em Computational Mechanics}, 57(6):1017--1035, 2016.

\bibitem{patil2018adaptive}
RU~Patil, BK~Mishra, and IV~Singh.
\newblock An adaptive multiscale phase field method for brittle fracture.
\newblock {\em Computer Methods in Applied Mechanics and Engineering}, 329:254--288, 2018.

\bibitem{giovanardi2017xfield}
Bianca Giovanardi, Anna Scotti, and Luca Formaggia.
\newblock A hybrid {XFEM}--phase field ({X}field) method for crack propagation in brittle elastic materials.
\newblock {\em Computer Methods in Applied Mechanics and Engineering}, 320:396--420, 2017.

\bibitem{geelen2020extended}
Rudy Geelen, Julia Plews, Michael Tupek, and John Dolbow.
\newblock An extended/generalized phase-field finite element method for crack growth with global-local enrichment.
\newblock {\em International Journal for Numerical Methods in Engineering}, 121(11):2534--2557, 2020.

\bibitem{muixi2021combined}
Alba Muix{\'\i}, Onofre Marco, Antonio Rodr{\'\i}guez-Ferran, and Sonia Fern{\'a}ndez-M{\'e}ndez.
\newblock A combined {XFEM} phase-field computational model for crack growth without remeshing.
\newblock {\em Computational Mechanics}, 67:231--249, 2021.

\bibitem{freddi2023adaptive}
Francesco Freddi and Lorenzo Mingazzi.
\newblock Adaptive mesh refinement for the phase field method: A fenics implementation.
\newblock {\em Applications in Engineering Science}, 14:100127, 2023.

\bibitem{bourdin2025variational}
Blaise Bourdin, Jean-Jacques Marigo, Corrado Maurini, and Camilla Zolesi.
\newblock A variational approach to fracture incorporating any convex strength criterion.
\newblock {\em arXiv preprint arXiv:2506.22558}, 2025.

\bibitem{vicentini2025variational}
Francesco Vicentini, Jonas Heinzmann, Pietro Carrara, and Laura De~Lorenzis.
\newblock Variational phase-field modeling of cohesive fracture with flexibly tunable strength surface.
\newblock {\em Journal of the Mechanics and Physics of Solids}, page 106424, 2025.

\bibitem{dolbow2025uniform}
Bo~Zeng, Johann Guilleminot, and John~E Dolbow.
\newblock Examining crack nucleation under spatially uniform stress states with a complete phase-field model for fracture.
\newblock {\em Theoretical and Applied Fracture Mechanics}, 140:105170, 2025.

\bibitem{chockalingam2025MCHB}
S~Chockalingam, Adrian~Buganza Tepole, and Aditya Kumar.
\newblock The phase-field model of fracture incorporating mohr-coulomb, mogi-coulomb, and hoek-brown strength surfaces.
\newblock {\em Engineering Fracture Mechanics}, 340:112108, 2026.

\bibitem{KKLP24}
Farhad Kamarei, Aditya Kumar, and Oscar Lopez-Pamies.
\newblock The poker-chip experiments of synthetic elastomers explained.
\newblock {\em Journal of the Mechanics and Physics of Solids}, 188:105683, 2024.

\bibitem{LP10}
Oscar Lopez-Pamies.
\newblock A new i1-based hyperelastic model for rubber elastic materials.
\newblock {\em Comptes Rendus Mecanique}, 338(1):3--11, 2010.

\bibitem{dahal2026dynamic}
Aarosh Dahal, Umar Khayaz, Ravindra Duddu, and Aditya Kumar.
\newblock Dynamic phase-field model for brittle fracture in grounded glaciers.
\newblock {\em arXiv preprint arXiv:2607.26274}, 2026.

\bibitem{liu2024effects}
Yangyuanchen Liu, Oscar Lopez-Pamies, and John~E Dolbow.
\newblock On the effects of material strength in dynamic fracture: A phase-field study.
\newblock {\em arXiv preprint arXiv:2411.16393}, 2024.

\bibitem{Hofacker2012_ContinuumPhaseField}
Martina Hofacker and Christian Miehe.
\newblock Continuum phase field modeling of dynamic fracture: variational principles and staggered fe implementation.
\newblock {\em International Journal of Fracture}, 178:113--129, 2012.

\bibitem{Nguyen2018_PhaseFieldCohesive}
Vinh~Phu Nguyen and Jian-Ying Wu.
\newblock Modeling dynamic fracture of solids with a phase-field regularized cohesive zone model.
\newblock {\em Computer Methods in Applied Mechanics and Engineering}, 340:1000--1022, 2018.
\newblock Available at: \url{www.elsevier.com/locate/cma}.

\bibitem{Geelen2019_PhaseField}
Rudy~J.M. Geelen, Yingjie Liu, Tianchen Hu, Michael~R. Tupek, and John~E. Dolbow.
\newblock A phase-field formulation for dynamic cohesive fracture.
\newblock {\em Computer Methods in Applied Mechanics and Engineering}, 348:680--711, 2019.
\newblock Available at: \url{www.elsevier.com/locate/cma}.

\bibitem{durussel2026dynamic}
Shad Durussel, Gergely Moln{\'a}r, and Jean-Fran{\c{c}}ois Molinari.
\newblock Origins of phase-field crack widening in dynamic fragmentation explained.
\newblock {\em Computer Methods in Applied Mechanics and Engineering}, 456:118942, 2026.

\bibitem{del2026dynamic}
Enrique~M del Castillo and Liuchi Li.
\newblock Assessing variational phase-field fracture against lefm predictions for crack-tip dynamics.
\newblock {\em International Journal of Solids and Structures}, 340:114199, 2026.

\bibitem{Lorenzis2026dynamic}
Jonas Heinzmann, Francesco Vicentini, Pietro Carrara, and Laura De~Lorenzis.
\newblock On phase-field regularization in dynamic fracture with brittle and cohesive formulations.
\newblock {\em arXiv preprint arXiv:2607.13599}, 2026.

\bibitem{zengdolbow2026thermal}
Bo~Zeng and John~E Dolbow.
\newblock A complete phase-field fracture model for brittle materials subjected to thermal shocks.
\newblock {\em arXiv preprint arXiv:2602.09031}, 2026.

\bibitem{freddi2022mesh}
Francesco Freddi and Lorenzo Mingazzi.
\newblock Mesh refinement procedures for the phase field approach to brittle fracture.
\newblock {\em Computer Methods in Applied Mechanics and Engineering}, 388:114214, 2022.

\bibitem{LK24}
Chang Liu and Aditya Kumar.
\newblock Emergence of tension–compression asymmetry from a complete phase-field approach to brittle fracture.
\newblock {\em International Journal of Solids and Structures}, 309:113170, 2025.

\bibitem{plaza2000mesh}
A~Plaza and GF~Carey.
\newblock Local refinement of simplicial grids based on the skeleton.
\newblock {\em Applied Numerical Mathematics}, 32(2):195--218, 2000.

\bibitem{manewaldscipy2020}
Pauli Virtanen, Ralf Gommers, Travis~E Oliphant, Matt Haberland, Tyler Reddy, David Cournapeau, Evgeni Burovski, Pearu Peterson, Warren Weckesser, Jonathan Bright, et~al.
\newblock {SciPy} 1.0: fundamental algorithms for scientific computing in {P}ython.
\newblock {\em Nature Methods}, 17(3):261--272, 2020.

\bibitem{kirk2006libmesh}
Benjamin~S Kirk, John~W Peterson, Roy~H Stogner, and Graham~F Carey.
\newblock lib{M}esh: a {C}++ library for parallel adaptive mesh refinement/coarsening simulations.
\newblock {\em Engineering with Computers}, 22(3-4):237--254, 2006.

\bibitem{arndt2021dealii}
Daniel Arndt, Wolfgang Bangerth, Denis Davydov, Timo Heister, Luca Heltai, Guido Kanschat, Martin Kronbichler, Matthias Maier, Jean-Paul Pelteret, Bruno Turcksin, and David Wells.
\newblock The deal.{II} library, version 9.3.
\newblock {\em Journal of Numerical Mathematics}, 29(3):171--186, 2021.

\bibitem{kim2024octree}
Ho-Young Kim and Hyun-Gyu Kim.
\newblock An adaptive continuous--discontinuous approach for the analysis of phase field fracture using mesh refinement and coarsening schemes and octree-based trimmed hexahedral meshes.
\newblock {\em Computational Mechanics}, 74:1171--1196, 2024.

\bibitem{bazantoh1983}
Zden{\v{e}}k~P Ba{\v{z}}ant and Byung~H Oh.
\newblock Crack band theory for fracture of concrete.
\newblock {\em Mat{\'e}riaux et construction}, 16:155--177, 1983.

\bibitem{stroblseelig2015}
Michael Strobl and Thomas Seelig.
\newblock A novel treatment of crack boundary conditions in phase field models of fracture.
\newblock {\em Pamm}, 15(1):155--156, 2015.

\bibitem{steinke2019}
Christian Steinke and Michael Kaliske.
\newblock A phase-field crack model based on directional stress decomposition.
\newblock {\em Computational Mechanics}, 63:1019--1046, 2019.

\bibitem{barrata2023dolfinx}
Igor~A Baratta, Joseph~P Dean, J{\o}rgen~S Dokken, Michal Habera, Jack~S Hale, Chris~N Richardson, Marie~E Rognes, Matthew~W Scroggs, Nathan Sime, and Garth~N Wells.
\newblock {DOLFIN}x: the next generation {FEniCS} problem solving environment.
\newblock {\em preprint}, 2023.

\bibitem{logg2012automated}
Anders Logg, Kent-Andre Mardal, and Garth Wells.
\newblock {\em Automated Solution of Differential Equations by the Finite Element Method: The {FEniCS} Book}, volume~84.
\newblock Springer Science \& Business Media, 2012.

\bibitem{CrouzeixRaviart73}
Michel Crouzeix and P-A Raviart.
\newblock Conforming and nonconforming finite element methods for solving the stationary stokes equations i.
\newblock {\em Revue fran{\c{c}}aise d'automatique informatique recherche op{\'e}rationnelle. Math{\'e}matique}, 7(R3):33--75, 1973.

\bibitem{balay2019petsc}
Satish Balay, Shrirang Abhyankar, Mark~F Adams, et~al.
\newblock {PETSc} users manual.
\newblock {\em Argonne National Laboratory}, 2019.

\bibitem{benson2006flexible}
Steven~J. Benson and Todd~S. Munson.
\newblock Flexible complementarity solvers for large-scale applications.
\newblock {\em Optimization Methods and Software}, 21(1):155--168, 2006.

\bibitem{amestoy2001mumps}
Patrick~R. Amestoy, Iain~S. Duff, Jean-Yves L'Excellent, and Jacko Koster.
\newblock A fully asynchronous multifrontal solver using distributed dynamic scheduling.
\newblock {\em SIAM Journal on Matrix Analysis and Applications}, 23(1):15--41, 2001.

\bibitem{henson2002boomeramg}
Van~Emden Henson and Ulrike~Meier Yang.
\newblock {BoomerAMG}: A parallel algebraic multigrid solver and preconditioner.
\newblock {\em Applied Numerical Mathematics}, 41(1):155--177, 2002.

\bibitem{Hilber1977_ImprovedDissipation}
Hans~M. Hilber, Thomas~J.R. Hughes, and Robert~L. Taylor.
\newblock Improved numerical dissipation for time integration algorithms in structural dynamics.
\newblock {\em Earthquake Engineering and Structural Dynamics}, 5:283--292, 1977.

\bibitem{hossain2014}
MZ~Hossain, C-J Hsueh, B~Bourdin, and K~Bhattacharya.
\newblock Effective toughness of heterogeneous media.
\newblock {\em Journal of the Mechanics and Physics of Solids}, 71:15--32, 2014.

\bibitem{KDK2025Comparison}
Umar Khayaz, Aarosh Dahal, and Aditya Kumar.
\newblock A comparison of phase field models of brittle fracture incorporating strength, {I}: {M}ixed-mode loading.
\newblock {\em Engineering Fracture Mechanics}, 330:111679, 2025.

\bibitem{lazarus2008comparison}
Veronique Lazarus, F-G Buchholz, M~Fulland, and J~Wiebesiek.
\newblock Comparison of predictions by mode {II} or mode {III} criteria on crack front twisting in three or four point bending experiments.
\newblock {\em International Journal of Fracture}, 153(2):141--151, 2008.

\bibitem{pham2017echelon}
KH~Pham and K~Ravi-Chandar.
\newblock Further examination of the criterion for crack initiation under mixed-mode {I}+ {III} loading.
\newblock {\em International Journal of Fracture}, 189(2):121--138, 2014.

\bibitem{chen2015crack3d}
C-H Chen, T~Cambonie, V~Lazarus, M~Nicoli, AJ~Pons, and A~Karma.
\newblock Crack front segmentation and facet coarsening in mixed-mode fracture.
\newblock {\em Physical Review Letters}, 115:265503, 2015.

\bibitem{kamarei2026nine}
Farhad Kamarei, Bo~Zeng, John~E. Dolbow, and Oscar Lopez-Pamies.
\newblock Nine circles of elastic brittle fracture: A series of challenge problems to assess fracture models.
\newblock {\em Computer Methods in Applied Mechanics and Engineering}, 448:118449, 2026.

\bibitem{Song2008_ComparativeFE}
Jeong-Hoon Song, Hongwu Wang, and Ted Belytschko.
\newblock A comparative study on finite element methods for dynamic fracture.
\newblock {\em Computational Mechanics}, 42:239--250, 2008.

\bibitem{Bleyer2017_DynamicCrackVariational}
Jérémy Bleyer, Clément Roux-Langlois, and Jean-François Molinari.
\newblock Dynamic crack propagation with a variational phase-field model: limiting speed, crack branching and velocity-toughening mechanisms.
\newblock {\em International Journal of Fracture}, 204:79--100, 2017.

\bibitem{shao2011quenching}
Yingfeng Shao, Yong Zhang, Xianghong Xu, Zhilong Zhou, Wei Li, and Bing Liu.
\newblock Effect of crack pattern on the residual strength of ceramics after quenching.
\newblock {\em Journal of the American Ceramic Society}, 94(9):2804--2807, 2011.

\bibitem{jiang2012thermal}
C.~P. Jiang, X.~F. Wu, J.~Li, F.~Song, Y.~F. Shao, X.~H. Xu, and P.~Yan.
\newblock A study of the mechanism of formation and numerical simulations of crack patterns in ceramics subjected to thermal shock.
\newblock {\em Acta Materialia}, 60(11):4540--4550, 2012.

\bibitem{sicsic2014thermal}
Paul Sicsic, Jean-Jacques Marigo, and Corrado Maurini.
\newblock Initiation of a periodic array of cracks in the thermal shock problem: A gradient damage modeling.
\newblock {\em Journal of the Mechanics and Physics of Solids}, 63:256--284, 2014.

\bibitem{anderson1965iterative}
Donald~G. Anderson.
\newblock Iterative procedures for nonlinear integral equations.
\newblock {\em Journal of the ACM}, 12(4):547--560, 1965.

\bibitem{walker2011anderson}
Homer~F. Walker and Peng Ni.
\newblock Anderson acceleration for fixed-point iterations.
\newblock {\em SIAM Journal on Numerical Analysis}, 49(4):1715--1735, 2011.

\bibitem{storvik2021accelerated}
Erlend Storvik, Jakub~Wiktor Both, Juan~Michael Sargado, Jan~Martin Nordbotten, and Florin~Adrian Radu.
\newblock An accelerated staggered scheme for variational phase-field models of brittle fracture.
\newblock {\em Computer Methods in Applied Mechanics and Engineering}, 381:113822, 2021.

\bibitem{rohracker2026efficient}
Maurice Rohracker, Paras Kumar, Paul Steinmann, and Julia Mergheim.
\newblock Efficient phase-field fracture simulations for fracture analysis in heterogeneous materials.
\newblock {\em Computational Mechanics}, 77(3):705--727, 2026.

\bibitem{waisman2026acceleration}
Shqipron Shala and Haim Waisman.
\newblock Acceleration scheme for nonlocal gradient damage methods.
\newblock {\em International Journal for Numerical Methods in Engineering}, 127(14):e70386, 2026.

\end{thebibliography}

\end{document}